\documentclass{article}

\usepackage{arxiv}

\usepackage[utf8]{inputenc} 
\usepackage[T1]{fontenc}    
\usepackage{hyperref}       
\usepackage{url}            
\usepackage{booktabs}       
\usepackage{amsfonts}       
\usepackage{nicefrac}       
\usepackage{microtype}      
\usepackage{lipsum}
\usepackage{graphicx}
\graphicspath{ {./images/} }

\usepackage{array}
\usepackage{pifont}
\usepackage{amsmath}
\newcolumntype{C}[1]{>{\centering\arraybackslash}p{#1}}

\newcommand{\cmark}{\text{\ding{51}}}
\newcommand{\xmark}{\text{\ding{55}}}

\title{Security and Privacy in Retrieval-Augmented Generation: Architectures, Threats, Defenses, and Future Directions for Building Trustworthy Systems}

\author{
 Balamurugan Palanisamy \\
  Department of Electrical and Electronics Engineering\\
  Birla Institute of Technology and Science, Pilani, Pilani Campus\\
  Vidya Vihar, Pilani, Rajasthan 333031, India \\
  \texttt{p20220455@pilani.bits-pilani.ac.in} \\
   \And
 G S S Chalapathi \\
  Department of Electrical and Electronics Engineering\\
  Birla Institute of Technology and Science, Pilani, Pilani Campus\\
  Vidya Vihar, Pilani, Rajasthan 333031, India \\
  \texttt{gssc@pilani.bits-pilani.ac.in} \\
  \And
 Vikas Hassija \\
  Department of Computer Engineering, KIIT University\\
  Bhubaneshwar, Odisha 751024, India \\
  \texttt{vikas.hassijafcs@kiit.ac.in} \\
  \And
 Rajkumar Buyya \\
  Quantum Cloud Computing and Distributed Systems (qCLOUDS) Laboratory \\ 
Department of Computing and Information Systems\\
  The University of Melbourne, Melbourne, Australia \\
   \texttt{rbuyya@unimelb.edu.au} \\
}

\begin{document}
\maketitle
\begin{abstract}
Retrieval-Augmented Generation (RAG) has emerged as a dominant paradigm for enhancing large language models with external knowledge. By coupling retrieval mechanisms with generative models, RAG systems improve factual grounding and adaptability across domains. However, integrating retrieval pipelines introduces new security and privacy risks that extend beyond conventional language modeling threats. Sensitive information may be exposed through retrieval indices, query logs, context construction, or federated updates, while adversarial manipulation of knowledge bases can undermine trust in generated outputs. This survey provides a comprehensive examination of privacy and security challenges across RAG systems deployed in centralized, on-device (Micro-RAG), federated, and hybrid paradigms. We present a unified taxonomy of threat surfaces spanning the retrieval, context construction, and generation stages and systematically analyze attack classes, including membership inference, index inference, poisoning, gradient leakage, and collusion. We further review architectural, algorithmic, and cryptographic defenses, highlighting privacy–utility trade-offs and deployment considerations. Finally, we outline open research challenges toward building trustworthy, secure, and resilient RAG systems for real-world applications.
\end{abstract}

\section{Introduction}
\label{sec:introduction}

Large Language Models (LLMs) have achieved strong performance across natural language processing tasks such as question answering, summarization, and decision support. However, their deployment in privacy-sensitive domains, including healthcare, finance, law, and enterprise knowledge management, remains challenging because centralized LLM services often require user queries, documents, or intermediate representations to be transmitted to cloud infrastructures \cite{llmsurvey1,llmsurvey2}. Retrieval-Augmented Generation (RAG) addresses some limitations of standalone LLMs by grounding generation in external knowledge sources. By retrieving relevant documents and incorporating them into the model context, RAG improves factual grounding, enables knowledge updates without full retraining, and supports domain adaptation \cite{ragsurvey2}.

Despite these benefits, conventional RAG systems are predominantly centralized: retrieval indices, document stores, and generative models are hosted in the cloud. Such designs assume reliable connectivity, abundant computation, and permissive data-sharing policies, which are often unrealistic in edge, mobile, IoT, and cross-organizational environments \cite{ragsurvey3}. To address these limitations, on-device RAG, often referred to as Micro-RAG, performs retrieval and sometimes generation locally on resource-constrained devices, while Federated RAG enables multiple clients or organizations to collaboratively improve retrieval or generation components without sharing raw data \cite{ondevicerag1,flsurvey4}. Hybrid edge--cloud RAG combines these approaches by selectively partitioning retrieval, context construction, and generation across local and cloud resources.

However, decentralizing RAG introduces new security and privacy risks. Sensitive information may leak through retrieval indices, embeddings, query logs, generated outputs, or federated updates. At the same time, adversaries may manipulate retrieved evidence through prompt injection, retrieval poisoning, context manipulation, model poisoning, Sybil attacks, or local index tampering. These risks are amplified in on-device and federated settings, where systems must operate under limited memory, computation, energy, communication bandwidth, and monitoring capability. Therefore, trustworthy RAG requires a deployment-aware analysis that jointly considers architecture, privacy, security, and evaluation.

 \begin{table}[ht]
\centering
\footnotesize  
\caption{List of Abbreviations}
\label{tab:abbreviations}
\begin{tabular}{|l|p{0.389\textwidth}|l|p{0.32\textwidth}|}
\hline
\centering\textbf{Abbrev.} & \centering\textbf{Full Form} & \centering\textbf{Abbrev.} & \centering\arraybackslash\textbf{Full Form} \\
\hline
AUC    & Area Under the ROC Curve                & LLM    & Large Language Model \\
BEIR   & Benchmarking IR                          & MIA    & Membership Inference Attack \\
Byzantine FL & Byzantine-Fault-Tolerant FL          & MPC    & Multi-Party Computation \\
CRAG   & Comprehensive RAG Benchmark              & nDCG   & Normalized Discounted Cumulative Gain \\
CVE    & Common Vulnerabilities and Exposures     & NLP    & Natural Language Processing \\
DP     & Differential Privacy                     & NQ     & Natural Questions \\
DRS    & Directional Relative Shift               & NPU    & Neural Processing Unit \\
EM     & Exact Match                              & PII    & Personally Identifiable Information \\
FedAvg & Federated Averaging                      & PIR    & Private Information Retrieval \\
FedRAG & Federated RAG                            & PPML   & Privacy-Preserving Machine Learning \\
FL     & Federated Learning                       & QA     & Question Answering \\
FPR    & False Positive Rate                      & RAG    & Retrieval-Augmented Generation \\
GDPR   & General Data Protection Regulation       & ROC    & Receiver Operating Characteristic \\
GPU    & Graphics Processing Unit                 & SE     & Searchable Encryption \\
HE     & Homomorphic Encryption                   & SGX    & Software Guard Extensions \\
HIPAA  & Health Insurance Portability and Accountability Act & SLM & Small Language Model \\
IID    & Independent and Identically Distributed  & SMPC   & Secure Multi-Party Computation \\
IoT    & Internet of Things                       & SPOF   & Single Point of Failure \\
IPI    & Indirect Prompt Injection                & SSE    & Searchable Symmetric Encryption \\
IR     & Information Retrieval                    & TEE    & Trusted Execution Environment \\
KB     & Knowledge Base                           & TPR    & True Positive Rate \\
KNN    & k-Nearest Neighbor                       & top-k  & Top-k Retrieved Documents \\
\hline
\end{tabular}
\end{table}

\begin{figure*}[ht]
\centering
\includegraphics[scale=0.40]{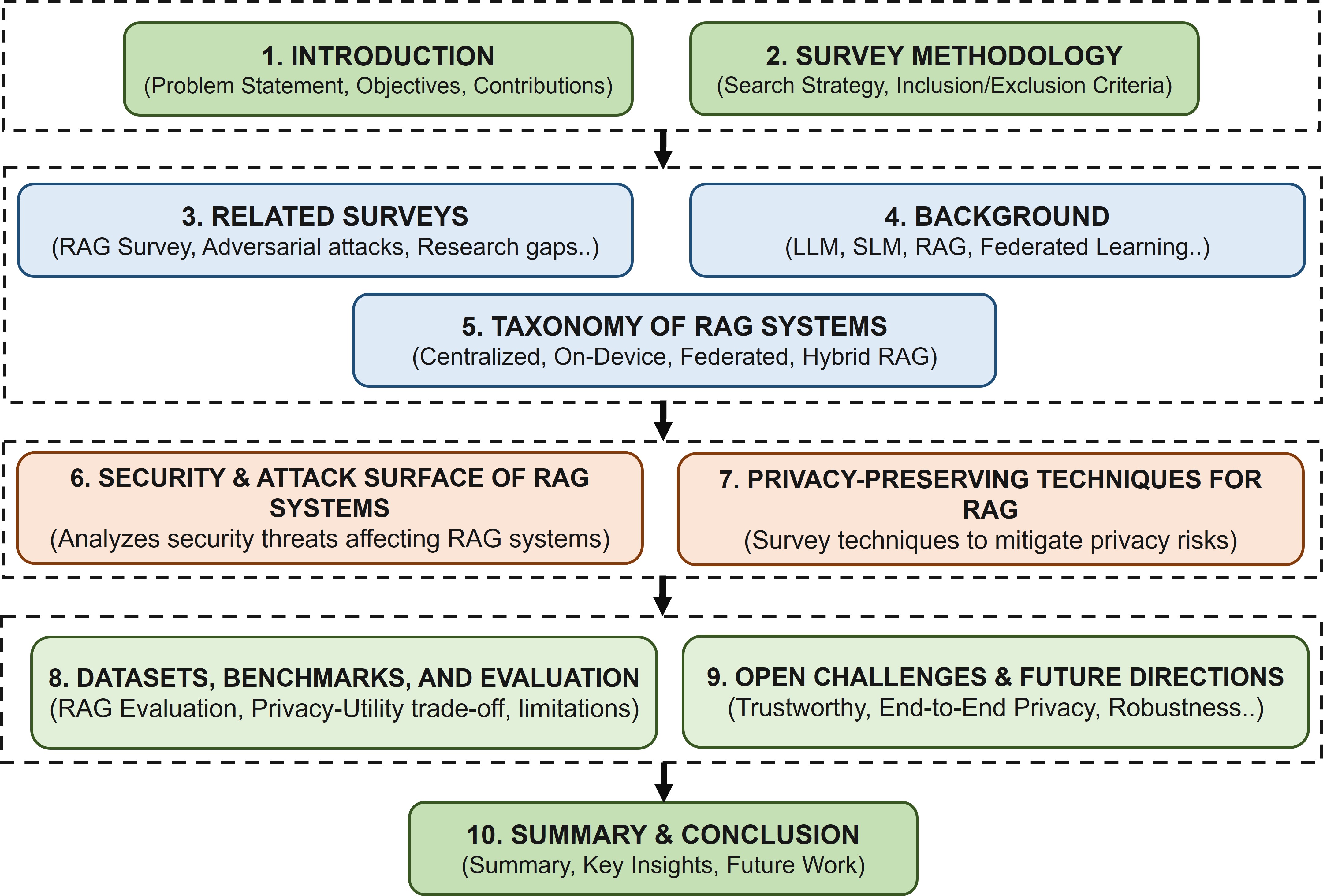}
\caption{Structural organization of the survey paper.}
\label{fig:paperorg}
\end{figure*}

Existing surveys have examined RAG architectures, LLM trustworthiness, adversarial attacks, or federated learning in isolation. However, a unified treatment of security and privacy across centralized, on-device, federated, and hybrid RAG systems remains limited. This survey fills this gap by providing a structured, deployment-aware analysis of RAG systems and their threat surfaces, defenses, evaluation practices, and open challenges. The key contributions are summarized as follows:

\begin{enumerate}
    \item \textbf{Unified taxonomy of RAG deployments:} We present a cross-paradigm taxonomy of centralized, on-device (Micro-RAG), federated, and hybrid edge--cloud RAG systems, highlighting architectural choices, data-residency assumptions, and deployment trade-offs.

    \item \textbf{Deployment-aware security and privacy analysis:} We analyze the expanded RAG attack surface across retrieval, context construction, generation, local device execution, and federated aggregation, covering prompt injection, poisoning, membership and index inference, local tampering, gradient leakage, and Sybil attacks.

    \item \textbf{Context-construction perspective:} We identify context construction and evidence packing as critical but underexplored vulnerability surfaces, showing how finite context budgets, ordering, truncation, and evidence displacement can affect robustness, privacy, and factual grounding.

    \item \textbf{Defense-in-depth framework:} We synthesize layered defenses across the RAG pipeline, including query filtering, retrieval protection, privacy-aware context assembly, generation verification, and system-level monitoring.

    \item \textbf{Evaluation landscape and research agenda:} We review benchmarks, datasets, and metrics for evaluating RAG systems from retrieval, generation, privacy, security, efficiency, and federated-learning perspectives, and identify open challenges for scalable, secure, and privacy-preserving RAG.
\end{enumerate}

As shown in Fig.~\ref{fig:paperorg}, the remainder of this paper is organized as follows. Section~\ref{sec:methodology} describes the survey methodology, Section~\ref{sec:relatedwork} reviews related surveys, and Section~\ref{sec:background} introduces the required background. Section~\ref{sec:taxonomy} presents the RAG deployment taxonomy. Sections~\ref{sec:securityandattack} and~\ref{sec:privacy} analyze security threats and privacy-preserving techniques, respectively. Section~\ref{sec:datasets} reviews benchmarks and evaluation methodologies, Section~\ref{sec:open_challenges} discusses open challenges, and Section~\ref{sec:conclusion} concludes the paper.

\section{Survey Methodology}
\label{sec:methodology}

This survey follows a structured literature review methodology to identify, classify, and synthesize research on security- and privacy-aware Retrieval-Augmented Generation (RAG), with emphasis on centralized, on-device, federated, and hybrid deployments. Relevant studies were collected from IEEE Xplore, ACM Digital Library, SpringerLink, ScienceDirect, ACL Anthology, arXiv, and Google Scholar. The search primarily covered publications from 2020 onward, corresponding to the emergence of modern RAG systems, while earlier works were included when they provided foundational concepts in federated learning, differential privacy, secure aggregation, encrypted retrieval, trusted execution environments, information retrieval, or adversarial machine learning. Search queries combined terms such as ``retrieval-augmented generation'', ``RAG security'', ``RAG privacy'', ``RAG poisoning'', ``prompt injection'', ``membership inference'', ``index inference'', ``on-device RAG'', ``Micro-RAG'', ``federated RAG'', ``encrypted retrieval'', ``secure aggregation'', ``differential privacy'', and ``RAG evaluation''.

Studies were included if they proposed, analyzed, or surveyed RAG architectures, retrieval pipelines, privacy-preserving mechanisms, adversarial threats, on-device or edge deployment, federated retrieval/generation, or evaluation methods relevant to RAG. Works were excluded if they focused solely on general LLM capabilities without relevance to retrieval or privacy/security, lacked sufficient technical detail, or did not contribute to the architectural, security, privacy, or evaluation dimensions considered in this survey. Because RAG security and federated RAG are rapidly evolving areas, recent preprints were retained when they addressed emerging RAG-specific problems not yet covered by archival publications.

The selected literature was organized along four dimensions: deployment paradigm, pipeline stage, attack type, and defense mechanism. Deployment paradigms include centralized RAG, on-device or Micro-RAG, federated RAG, and hybrid edge--cloud RAG. Pipeline stages include query processing, retrieval and indexing, context construction, generation, training or aggregation, and system monitoring. Security threats were grouped into prompt-based attacks, retrieval poisoning, membership and index inference, retriever manipulation, context manipulation, local device attacks, gradient leakage, Sybil attacks, and compound cross-layer attacks. Privacy-preserving techniques were classified as architectural isolation, algorithmic perturbation, cryptographic protection, hardware-assisted isolation, and pipeline-stage controls. The reviewed works were then synthesized qualitatively to compare representative approaches, identify coverage gaps, and analyze the relationship between deployment choices, threat models, defense mechanisms, privacy--utility trade-offs, and evaluation metrics.

This review is limited by the rapidly evolving nature of RAG research and by inconsistent terminology across the literature. Terms such as local RAG, edge RAG, Micro-RAG, federated retrieval, and federated RAG are sometimes used differently across studies. To address this, the survey adopts a unified taxonomy and maps related works into common architectural, security, privacy, and evaluation categories.

\section{Related Surveys}
\label{sec:relatedwork}

Existing literature relevant to this survey spans five overlapping threads: RAG architectures, federated learning, adversarial robustness, privacy-preserving retrieval, and edge/on-device AI. General RAG surveys by Fan \textit{et al.} \cite{intro3} and Sharma and Bhattarai \cite{centralrag1} provide broad taxonomies of retrieval mechanisms, pipeline designs, and application domains, while Li \textit{et al.} \cite{ragsurvey1} extend this discussion toward RAG-reasoning systems involving chain-of-thought and agentic workflows. Evaluation-oriented efforts such as CRAG \cite{cragbenchmark} and Know-Your-RAG \cite{ragsurvey3} further advance benchmark design for factual consistency and retrieval coverage. However, these works primarily emphasize functional RAG capabilities and evaluation, with limited treatment of deployment-specific security and privacy risks in on-device, federated, or hybrid settings.

A second body of work studies federated learning and its security implications. McMahan \textit{et al.} \cite{floriginalpaper} introduced federated learning for collaborative model training without raw data sharing, and subsequent surveys by Li \textit{et al.} \cite{flsurvey2} and Nguyen \textit{et al.} \cite{flsurvey1} examine challenges such as communication efficiency, non-IID data, resource constraints, and Byzantine robustness. Related studies have also investigated model poisoning, gradient leakage, and Sybil attacks, including Sybil-aware defenses \cite{collusion3}. Recent work on federated search for RAG \cite{limitation5} begins to connect FL with retrieval-augmented systems, but does not fully analyze the compound threat surface created when retrieval encoders, local knowledge indices, and generation models interact across distributed clients.

Security-focused studies provide another important foundation. Surveys on adversarial attacks against LLMs cover prompt injection, jailbreaking, and alignment subversion \cite{promptattack1,jailbreaking6}, while RAG-specific studies examine retrieval-stage leakage \cite{indexattack3}, enterprise RAG attack surfaces \cite{security1}, multi-stage adversarial evaluation through SafeRAG \cite{saferag}, and security countermeasures for RAG systems \cite{microrag1}. Individual attack frameworks such as PoisonedRAG \cite{poisoningattack2}, poisoning traceback \cite{poisoningattack1}, and BadRAG \cite{microrag2} have advanced understanding of retrieval poisoning and on-device vulnerabilities. Nevertheless, these works largely focus on specific attacks or deployment contexts rather than providing a unified cross-paradigm mapping of threats across centralized, on-device, federated, and hybrid RAG systems.

Privacy-preserving techniques for RAG draw from differential privacy, encrypted retrieval, trusted execution environments, and secure multi-party computation. Differential privacy has been applied to model updates, retrieval scores, and embedding representations \cite{limitation41}, while searchable encryption and homomorphic encryption support privacy-preserving retrieval with significant scalability constraints. Privacy-aware RAG for collaborative organizational settings has also been studied \cite{flsurvey5}. Similarly, TEEs and SMPC provide strong protection for confidential computation and federated aggregation, but their integration with high-dimensional vector retrieval remains challenging \cite{tees}. These approaches are usually studied as isolated mechanisms rather than as composable defenses across the full RAG pipeline.

Finally, edge AI and on-device inference surveys address resource constraints, model compression, quantization, and neural acceleration for embedded systems \cite{ondeviceai3,ondeviceai1}. Early Micro-RAG systems such as MeMemo \cite{ondevicerag1} and cloud-device collaborative personalization approaches \cite{ondevicerag2} demonstrate the feasibility of local retrieval and generation. However, existing edge AI literature rarely treats local knowledge indexing, retrieval tampering, side-channel leakage, and reduced monitoring capability as first-class security and privacy concerns.

Table \ref{tab:comparison} summarizes how representative surveys cover architectural taxonomy, security analysis, privacy mechanisms, on-device deployment, and federated coordination. In contrast to prior works, this survey jointly examines these dimensions and provides a deployment-aware treatment of RAG security and privacy across centralized, on-device, federated, and hybrid edge--cloud paradigms.

\begin{table*}[hb]
\centering
\footnotesize
\caption{Coverage comparison of representative related surveys. \cmark = full coverage; Partial = limited coverage; \xmark = not addressed.}
    \begin{tabular}{|m{3.9cm}|>{\centering\arraybackslash}m{2cm}|>{\centering\arraybackslash}m{2cm}|>{\centering\arraybackslash}m{2cm}|>{\centering\arraybackslash}m{2cm}|>{\centering\arraybackslash}m{2cm}|}
    \hline        
    \centering\arraybackslash\textbf{Survey / Work}&\centering\arraybackslash\textbf{RAG Taxonomy}&\centering\arraybackslash\textbf{Security Analysis} &\centering\arraybackslash\textbf{Privacy Mechanisms} &\centering\arraybackslash\textbf{On-Device / Edge} &\centering\arraybackslash\textbf{Federated Coordination}  \\ \hline

    Fan \textit{et al.} \cite{intro3} (2024) & \cmark & \xmark & \xmark & \cmark & \xmark \\  \hline
    
    Zeng \textit{et al.} \cite{indexattack3} (2024) & Partial & \cmark & \cmark & \xmark & \xmark \\  \hline

    Ni \textit{et al.} \cite{limitation71} (2025) & Partial & Partial & Partial & \xmark & \xmark \\ \hline

    Vonderhaar \textit{et al.} \cite{security1} (2025) & \xmark & \cmark & Partial & \xmark & \xmark \\  \hline

    Wang \textit{et al.} \cite{microrag1} (2025) & Partial & \cmark & Partial & \xmark & \xmark \\  \hline
	
    Chakraborty \textit{et al.} \cite{federatedrag1} (2025) & \cmark & Partial & Partial & \xmark & \cmark \\  \hline
	
    Sharma \textit{et al.} \cite{centralrag1} (2026) & \cmark & Partial & \xmark & \xmark & \xmark \\  \hline

    \textbf{This Survey} & \cmark & \cmark & \cmark & \cmark & \cmark \\  \hline

    \end{tabular}
    \label{tab:comparison} 
\end{table*}

\section{Background}
\label{sec:background}

This section summarizes the core concepts required for understanding secure and privacy-aware RAG deployments, including language models, retrieval-augmented generation, on-device constraints, and federated learning.

\subsection{Language Models and Retrieval-Augmented Generation}

Large Language Models (LLMs), typically based on Transformer architectures, have demonstrated strong generative and reasoning capabilities across tasks such as question answering, summarization, and dialogue. However, their large memory, compute, and energy requirements often favor centralized cloud deployment \cite{llmsurvey1,llmsurvey2}. Small Language Models (SLMs) provide lightweight alternatives for edge and embedded environments through compression techniques such as distillation, pruning, quantization, and architectural simplification \cite{slmsurvey1}. Although SLMs reduce deployment cost, they generally possess weaker world knowledge and reasoning capacity than LLMs, motivating the use of external knowledge through retrieval \cite{slmvsllm}.

\begin{figure*}[ht]
\centering
\includegraphics[width=1\textwidth]{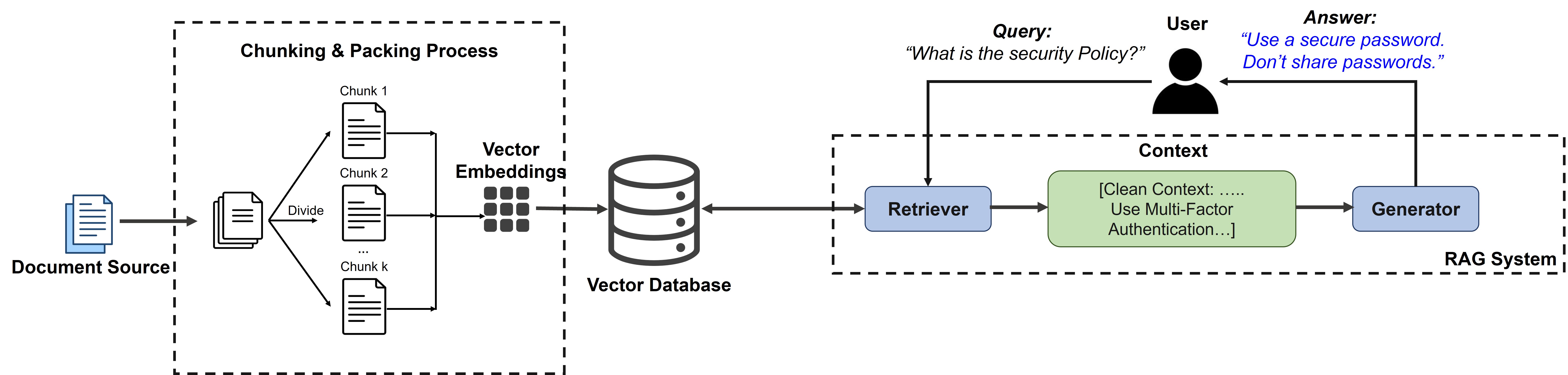}
\caption{RAG Pipeline: RAG system architecture showing offline indexing and online inference phases. During indexing, documents are chunked, embedded, and stored in a vector database. During inference, a query follows the retrieval--context--generation flow.}
\label{fig:ragpipeline}
\end{figure*}

Retrieval-Augmented Generation (RAG) grounds language model outputs in external knowledge rather than relying solely on parametric memory. A typical RAG pipeline, shown in Fig.~\ref{fig:ragpipeline}, consists of query formulation, document retrieval, context construction, and conditioned generation. By decoupling knowledge storage from model parameters, RAG improves factual grounding, supports knowledge updates through index refresh, reduces the need for full model retraining, and enables domain adaptation \cite{ragsurvey2}. Early RAG systems were largely cloud-based, but privacy-sensitive and resource-constrained applications increasingly challenge this centralized assumption \cite{ragsurvey3}.

\subsection{On-Device and Federated Deployment Constraints}

On-device RAG, or Micro-RAG, moves retrieval and sometimes generation to edge devices. This improves data locality, latency, and offline availability, but introduces strict limits on memory, computation, storage, and energy \cite{ondeviceai2}. Consequently, Micro-RAG systems often rely on compact vector indices, quantized or distilled models, selective retrieval, retrieval-only local processing, and partial cloud offloading. These design choices create a tension between privacy, efficiency, and retrieval coverage.

Federated Learning (FL) enables multiple clients to collaboratively train or adapt models without sharing raw data. In server-based FL, a central coordinator aggregates client updates, whereas decentralized or gossip-based FL relies on peer-to-peer exchange \cite{flsurvey3}. FL is attractive for privacy-sensitive RAG because clients can adapt retrievers, share statistical signals, or fine-tune components without centralizing local corpora. However, FL also introduces communication overhead, non-IID data distributions, heterogeneous client capabilities, unreliable participation, and adversarial risks.

Federated RAG combines these ideas by distributing retrieval and generation components across clients while preserving data sovereignty. Common mechanisms include distributed retriever adaptation, privacy-preserving representation learning, collective indexing or sketch sharing, and federated generator tuning \cite{flsurvey4}. Together, on-device and federated RAG shift the design problem from purely improving retrieval and generation quality toward balancing privacy, robustness, resource constraints, and collaborative intelligence.

\section{Taxonomy of Retrieval-Augmented Generation Systems}
\label{sec:taxonomy}

\begin{figure*}[hb]
\centering
\includegraphics[scale=0.35]{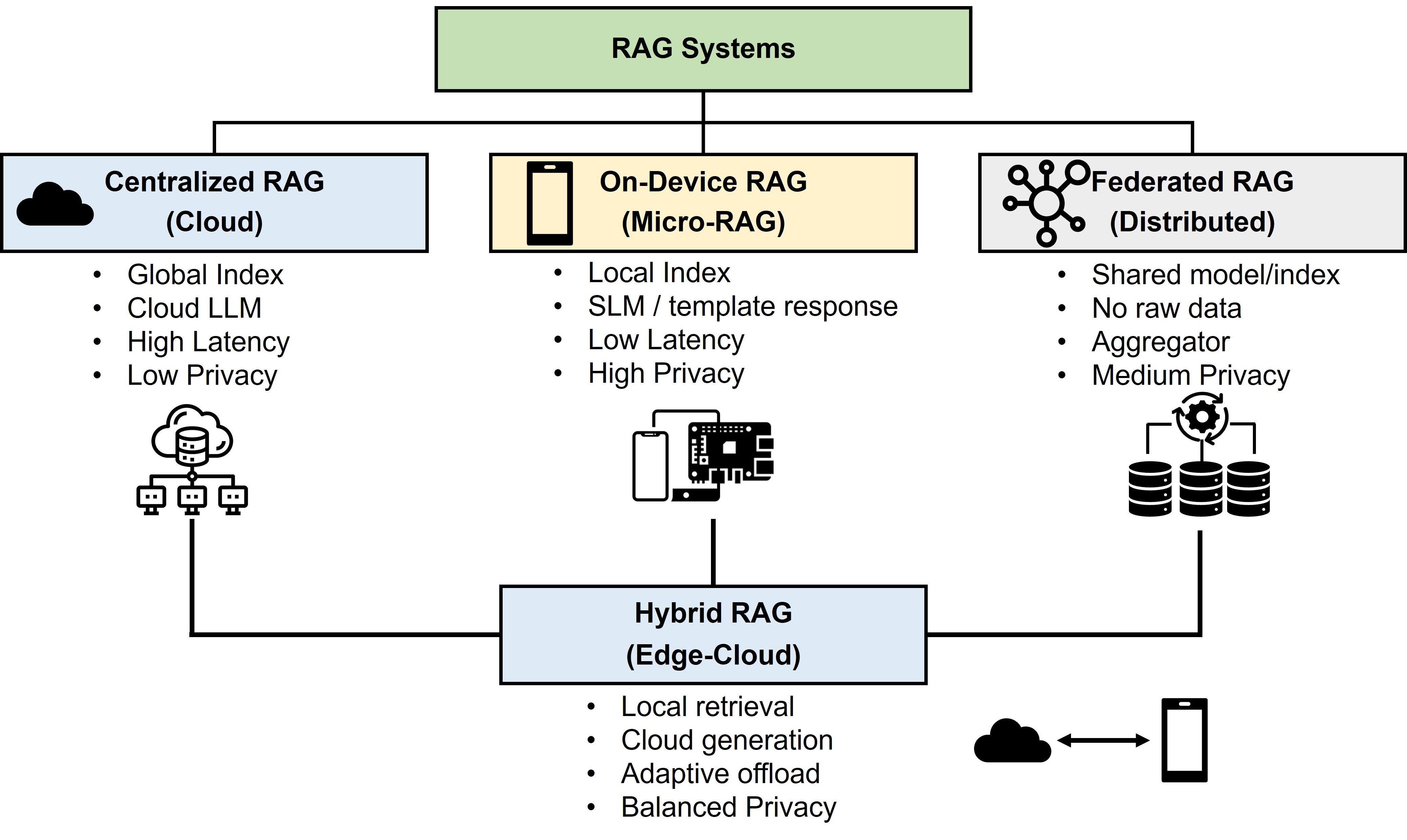}
\caption{Structural Taxonomy of RAG Systems: Categorizing Centralized, Micro-RAG, Federated, and Hybrid Approaches.}
\label{fig:ragtaxonomy}
\end{figure*}

This section classifies RAG systems according to four deployment dimensions: where data resides, where retrieval is performed, where generation is executed, and how knowledge or model updates are shared. Based on these dimensions, we group RAG systems into four major paradigms, as illustrated in Fig.~\ref{fig:ragtaxonomy}: centralized RAG, on-device or Micro-RAG, federated RAG, and hybrid edge--cloud RAG. This taxonomy is useful because deployment choices directly shape privacy guarantees, latency, scalability, resource requirements, and attack surfaces.

\subsection{Centralized RAG}

Centralized RAG is the most common deployment model for large-scale RAG services. In this paradigm, documents, retrieval indices, embedding models, and generative models are hosted within a cloud infrastructure. A user query is sent to the server, encoded, matched against a global knowledge base, and augmented with the top-$k$ retrieved passages before being passed to a cloud-hosted generator. This design benefits from high model capacity, global retrieval coverage, centralized maintenance, and efficient index updates \cite{centralrag2}.

However, centralized RAG also introduces important limitations. User queries, retrieved contexts, or latent representations may be exposed to the service provider, creating privacy and compliance concerns in domains such as healthcare, finance, law, and enterprise knowledge management. Centralized indices also create single points of failure, increase dependence on network connectivity, and may fail to exploit private or domain-specific data that cannot be uploaded to the cloud \cite{ragbottlenecks}. Thus, centralized RAG is well suited for open-domain or low-sensitivity applications, but less suitable when strict data locality, low latency, or organizational data sovereignty is required.

\subsection{On-Device or Micro-RAG}

On-device RAG, often referred to as Micro-RAG, moves retrieval and, in some cases, generation to local devices such as smartphones, embedded systems, industrial controllers, or IoT nodes. Its main objective is to preserve data locality by keeping user queries and private knowledge bases on the device. Micro-RAG commonly appears in three forms: fully local retrieval and lightweight generation, local retrieval with cloud-based generation, and local retrieval with templated or structured responses \cite{ondmicrorag1}.

This paradigm provides strong privacy, offline availability, reduced network dependence, and predictable latency. It is particularly relevant for personal assistants, clinical edge devices, industrial diagnostics, defense systems, and mobile IoT applications. Nevertheless, Micro-RAG is constrained by limited memory, storage, computation, and energy. These constraints restrict index size, retrieval depth, embedding dimensionality, and generator capacity. Practical systems therefore rely on compact vector indices, quantized or distilled models, selective retrieval, shallow retrievers, and partial offloading. The key design challenge is to balance privacy and responsiveness against retrieval coverage and response quality.

\subsection{Federated RAG}

Federated RAG extends RAG to multi-client or cross-silo environments where data cannot be centrally pooled due to privacy, legal, organizational, or competitive constraints. Instead of sharing raw documents or queries, clients may collaboratively adapt retrievers, align embedding spaces, share aggregate statistics, or fine-tune generation components through federated learning mechanisms. Coordination may be server-led, using a central aggregator such as FedAvg, or decentralized, using peer-to-peer or gossip-based communication.

FedRAG enables collaborative intelligence across distributed knowledge silos while preserving local data ownership. This makes it attractive for multi-hospital clinical systems, financial institutions, legal organizations, cybersecurity monitoring, and enterprise knowledge networks. However, it also introduces significant system and security challenges. Communication overhead, client heterogeneity, non-IID corpora, stragglers, version inconsistency, and malicious client behavior can degrade retrieval and generation quality. In addition, synchronizing retrieval encoders, local indices, and generators is more complex than standard federated model training because retrieval quality depends not only on model parameters but also on the structure and freshness of distributed knowledge bases.

\subsection{Hybrid Edge--Cloud RAG}

Hybrid RAG partitions retrieval, context construction, and generation across local and cloud resources. It is designed to balance the privacy of local processing with the reasoning capacity and scalability of cloud-hosted models. Common patterns include edge-side pre-filtering followed by cloud retrieval, local retrieval followed by cloud generation, and dynamic offloading based on query sensitivity, latency, battery level, or local compute availability.

Hybrid RAG offers a practical middle ground when fully local execution is too resource-intensive and full centralization is not acceptable. It can reduce bandwidth usage, preserve sensitive documents locally, and provide graceful degradation when connectivity is limited. However, hybrid designs introduce additional orchestration complexity and expand the attack surface across edge--cloud trust boundaries. Sensitive snippets, query embeddings, routing metadata, or partially processed context may cross system boundaries, requiring careful policy design, authentication, encryption, and selective disclosure.

\begin{table}[ht]
\footnotesize
    \centering
        \caption{Multi-Dimensional Comparison of RAG Paradigms}

    \begin{tabular}{|p{2.8cm}|p{2.8cm}|p{2.6cm}|p{2.8cm}|p{2.6cm}|}
        \hline
       \centering\textbf{Dimension} & \centering\textbf{Centralized RAG} & \centering\textbf{On-Device (Micro-RAG)}  & \centering\textbf{Federated RAG} & \centering\arraybackslash\textbf{Hybrid Edge–Cloud RAG} \\ \hline

        Data Residency & Centralized Cloud & Local Device & Distributed Silos & Partitioned (Edge/Cloud) \\ \hline

        Retrieval Location & Cloud & Device & Distributed & Device + Cloud \\ \hline

        Retrieval Scope & Global Knowledge & Local Context Only & Federated Collective & Tiered (Local + Global) \\ \hline

        Generation Location & Cloud & Device / Partial & Distributed / Central & Device + Cloud \\ 

        Privacy Level & Low–Medium & High & High & Medium–High \\ \hline

        Primary Privacy Guarantee & Encryption at Rest/Transit & Data Isolation & Local Processing & Selective Disclosure \\ \hline

        Communication Cost & High & Low & Medium–High & Medium \\ \hline

        Inference Latency & High (Network Dependent) & Deterministic (Local) & Synchronization Dependent & Adaptive \\ \hline

        Model Capacity & High (State-of-the-Art) & Low (Distilled/SLM) & Variable (Collaborative) & High (Offloaded) \\ \hline

        Implementation Complexity & Low (Established) & Medium (Optimization) & High (Orchestration) & Medium (Policy Design) \\ \hline

        Deployment Complexity & Low & Medium & High & Medium-High \\ \hline

    \end{tabular}
    \label{tab:ragcomparision}
\end{table}

\subsection{Comparative Analysis}

Table~\ref{tab:ragcomparision} summarizes the main differences among the four paradigms. No single architecture dominates across all dimensions. Centralized RAG provides the highest model capacity and operational simplicity, but it weakens data locality. Micro-RAG maximizes privacy and offline operation, but is limited by device resources and local knowledge coverage. Federated RAG enables collaboration across distributed silos, but introduces communication, heterogeneity, and trust-management challenges. Hybrid RAG offers a flexible compromise, but requires careful orchestration across multiple trust boundaries.

\begin{figure*}[hb]
\centering
\includegraphics[scale=0.35]{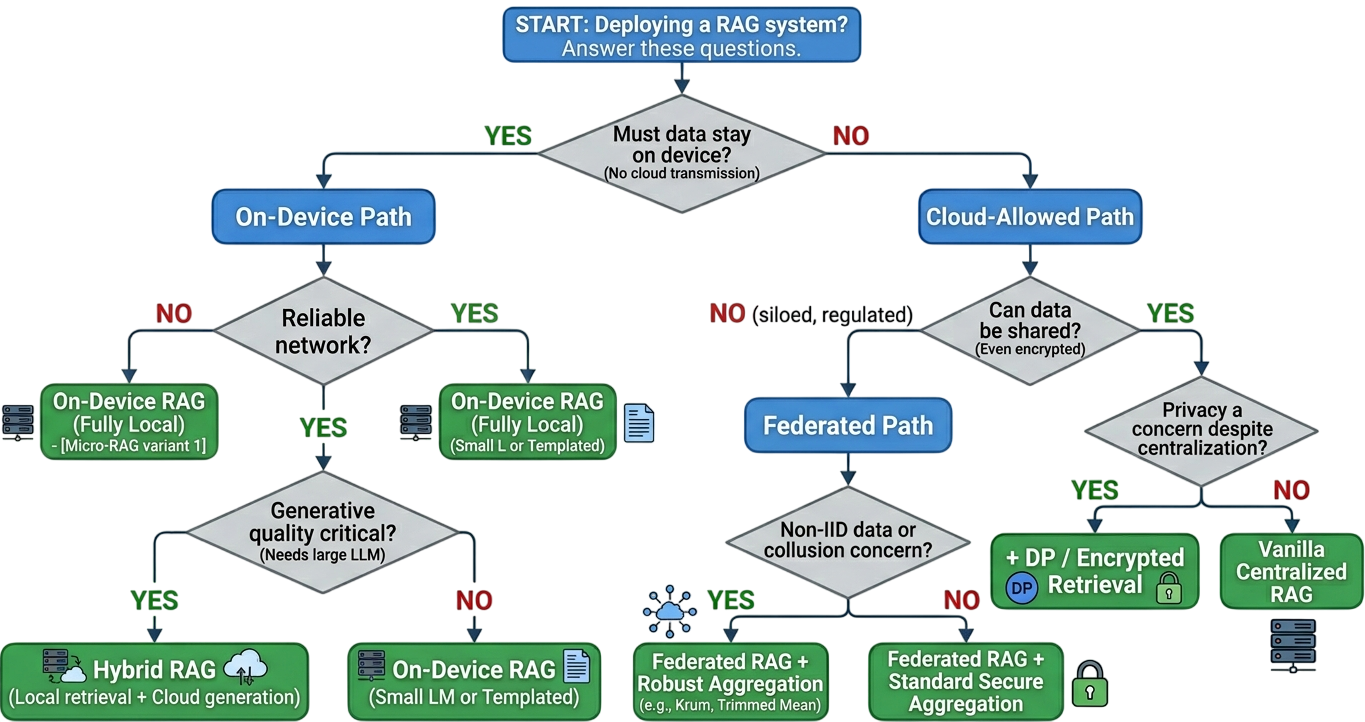}
\caption{RAG paradigm selection flowchart. Guides practitioners from deployment constraints (data locality, connectivity, compute budget, collaboration needs) to the optimal architecture among centralized, on-device, federated, or hybrid RAG, including defense recommendations.}
\label{fig:ragflowchart}
\end{figure*}

Overall, the choice of RAG architecture should be guided by application requirements such as data sensitivity, connectivity, latency, compute budget, collaboration needs, and acceptable privacy--utility trade-offs. Fig.~\ref{fig:ragflowchart} provides a decision-oriented view of these trade-offs. This deployment taxonomy provides the basis for the security, privacy, and evaluation analysis in the following sections.

\section{Security and Attack Surface of RAG Systems}
\label{sec:securityandattack}

\begin{figure*}[ht]
\centering
\includegraphics[width=1\textwidth]{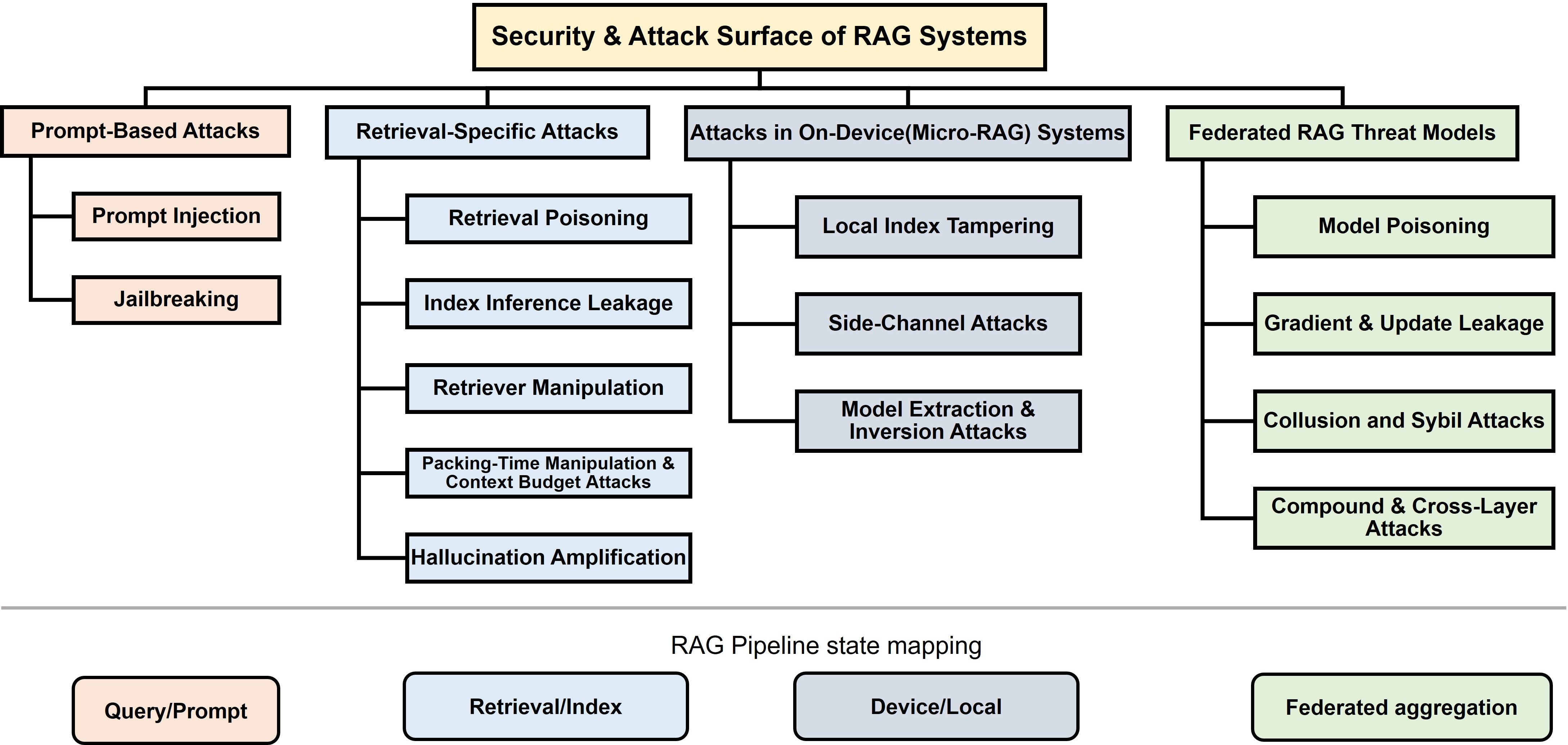}
\caption{Taxonomy of security threats and attack surfaces in RAG systems. Attacks are organized by deployment layer and pipeline stage.}
\label{fig:ragsecurityattacks}
\end{figure*}

RAG systems expand the attack surface of standalone language models by introducing external knowledge stores, retrievers, vector indices, context-construction policies, and, in decentralized settings, local devices or federated clients. Security failures may therefore arise not only from the generator but also from the retrieval corpus, the embedding model, the context window, device storage, or the federated aggregation process. This section organizes RAG threats by pipeline stage and deployment paradigm. The taxonomy in Fig.~\ref{fig:ragsecurityattacks} summarizes how these threats map to the RAG pipeline and system architecture.

\subsection{Pipeline-Level Threats}

\subsubsection{Prompt-Based Attacks}

Prompt-based attacks exploit the sensitivity of language models to input instructions. In RAG systems, this risk is amplified because retrieved documents are automatically inserted into the prompt and may be interpreted by the model as trusted context. Prompt injection attacks can be direct, where the user explicitly attempts to override system instructions, or indirect, where malicious instructions are embedded inside retrieved documents, web pages, logs, or external knowledge sources \cite{promptattack2,promptattack3}. Indirect Prompt Injection (IPI) is particularly relevant to RAG because the adversarial instruction enters the model through the retrieval pipeline rather than through the user query.

Successful prompt injection can lead to instruction override, system prompt leakage, output manipulation, or unauthorized tool use in agentic RAG systems \cite{promptattack4}. Jailbreaking is closely related but focuses more directly on bypassing safety alignment through adversarial framing, instruction layering, or context-induced alignment subversion \cite{jailbreaking1,jailbreaking3}. In RAG settings, retrieved documents can provide an apparently grounded justification for unsafe or policy-violating responses, weakening the separation between trusted system instructions and untrusted external content \cite{jailbreaking4,jailbreaking5}.

Mitigations include strict separation between instructions and retrieved content, input and output filtering, adversarially robust prompting, alignment-aware retrieval, and post-generation safety verification \cite{promptattack5,jailbreaking8}. However, these defenses are not sufficient in isolation because RAG systems continuously ingest dynamic and potentially untrusted external content. Prompt-based attacks therefore require defenses across retrieval, context construction, and generation rather than only at the prompt level.

\subsubsection{Retrieval and Index Attacks}

Retrieval-specific attacks target the non-parametric memory of RAG systems, including the document corpus, vector index, retriever, and ranking function. Retrieval poisoning inserts adversarial or misleading documents into the corpus so that they are retrieved for target queries and included in the generation context \cite{poisoningattack2}. Such attacks may use semantic anchoring, adversarial suffixes, embedding-space manipulation, or shadow retrievers to increase the probability that poisoned documents appear in the top-$k$ results \cite{poisoningattack3,poisoningattack6}. Unlike one-time prompt attacks, poisoning creates persistent vulnerabilities because the corrupted evidence remains in the knowledge base until detected and removed.

Index inference and membership inference attacks exploit observable retrieval behavior to infer sensitive information about the indexed corpus \cite{indexattack1,indexattack3}. Even when raw documents are not exposed, ranked outputs, similarity scores, citations, retrieval success rates, or repeated response patterns may reveal whether specific documents, topics, entities, or relationships are present in the index \cite{indexattack2,indexattack4}. These risks are especially relevant in enterprise, healthcare, legal, and federated RAG settings, where the index itself may encode confidential organizational knowledge.

Retriever manipulation attacks compromise the embedding or ranking model rather than the corpus. A manipulated retriever can systematically redirect classes of queries toward adversarial or irrelevant documents even when the index is clean \cite{microrag3}. This may occur through backdoored retrievers, embedding-space distortion, or malicious model updates in federated settings \cite{microrag2}. Since the generator can only reason over retrieved context, a compromised retriever undermines the entire RAG pipeline.

Defenses against retrieval and index attacks include corpus provenance checks, vector-space anomaly detection, retriever integrity testing, ranking-distribution monitoring, rate limiting, metadata suppression, randomized or coarsened retrieval outputs, differential privacy, and encrypted retrieval \cite{poisoningattack9,limitation31}. These mechanisms involve trade-offs: stronger privacy or robustness may reduce retrieval accuracy, increase latency, or complicate deployment, especially in on-device and federated environments.

\subsubsection{Context-Construction and Packing Attacks}

Context construction is a critical but underexplored attack surface. After retrieval, the system must select, order, truncate, and pack evidence into a finite context window. These decisions determine what information is visible to the generator. Context manipulation attacks exploit this stage by causing misleading or strategically framed documents to dominate the prompt, even when relevant evidence exists elsewhere in the index \cite{contextattack1}. Common strategies include length domination, semantic saturation, instruction overshadowing, and ranking exploitation.

Packing-time manipulation and context-budget attacks operate at the boundary between retrieval and generation. An adversary can exploit fixed token budgets, naive first-fit packing, truncation boundaries, or positional biases such as the ``lost-in-the-middle'' effect to displace legitimate evidence or place misleading content in highly attended positions \cite{promptattack7,addref3}. These attacks are difficult to detect because the individual documents may appear benign and relevant when evaluated in isolation. The adversarial effect emerges from the interaction between retrieved documents and the packing policy.

Mitigations require structural changes to context assembly, including diversity-constrained packing, source balancing, semantic-boundary truncation, adaptive summarization, evidence attribution, and faithfulness verification \cite{contextattack3,addref2}. For resource-constrained RAG systems, these defenses must be lightweight because excessive verification can negate the latency and energy advantages of local or edge deployment.

\subsubsection{Generation-Stage Reliability Attacks}

Although RAG is designed to reduce hallucination, retrieval can also amplify hallucination when the supplied context is incomplete, conflicting, outdated, or adversarially crafted. Hallucination amplification occurs when the generator produces unsupported or false claims with higher confidence because misleading retrieved evidence appears authoritative \cite{addref4}. This can arise from conflicting evidence, out-of-scope retrieval, over-reliance on retrieved content, or adversarially authoritative writing styles \cite{addref1}. The risk is particularly severe in high-stakes domains where users may interpret retrieved citations as proof of correctness.

Generation-stage defenses include consistency-based retrieval filtering, conflict detection, calibrated abstention, self-consistency checks, attribution-aware generation, and adversarial robustness training \cite{promptattack6,limitation32}. These mechanisms should be treated as part of a full-pipeline security strategy: the generator cannot reliably compensate for missing, poisoned, or badly packed evidence unless retrieval and context construction are also monitored.

\subsection{Deployment-Specific Threats}

\subsubsection{On-Device and Micro-RAG Threats}

On-device RAG improves privacy by keeping queries, documents, and retrieval operations local, but it shifts the threat model toward local compromise and physical access. Micro-RAG systems often store compact indices and compressed models on devices that may be lost, stolen, infected with malware, or operated without continuous monitoring. As a result, local index tampering becomes a major risk: an adversary may inject misleading documents, modify embeddings, or delete critical evidence to bias retrieval outcomes.

On-device RAG is also exposed to side-channel attacks, where timing, memory access, cache behavior, or power consumption reveals information about queries, retrieved documents, or index structure \cite{sidechannel1,sidechannel3}. In addition, compressed or distilled models used in Micro-RAG may be susceptible to model extraction or inversion attacks through repeated querying \cite{modelinv1}. These attacks can reveal properties of the model, local corpus, or retrieval behavior.

The central tension in Micro-RAG is privacy through isolation versus security through oversight. Local processing reduces server-side leakage but limits centralized logging, anomaly detection, and coordinated defense. Practical defenses include tamper-resistant storage, encrypted local indices, lightweight attestation, sandboxing, periodic integrity checks, and resource-aware context verification.

\begin{table*}[b]
\footnotesize
    \centering
        \caption{Cross-Paradigm Security Comparison for RAG Systems}
    \begin{tabular}{|p{2.2cm}|p{2.8cm}|p{2.8cm}|p{2.6cm}|p{2.8cm}|}
        \hline
\centering\textbf{Security Aspect} & \centering\textbf{Centralized RAG} & \centering\textbf{On-Device (Micro-RAG)}  & \centering\textbf{Federated RAG} & \centering\arraybackslash\textbf{Hybrid Edge–Cloud RAG} \\ \hline

        Dominant Threats & Prompt injection, retrieval poisoning & Local tampering, side channels & Poisoning, leakage, Sybil attacks & Combined edge + cloud threats \\ \hline

        Dominant Threat Vectors & Mass Prompt Injection, Index Poisoning & Index Tampering, Model Extraction & Client Collusion, Update Leakage &  Multi-tier Compromise, Policy Drift \\ \hline

        Primary Attack Surface & Cloud Infrastructure \& Global API & Physical Device \& Local Memory & Client-to-Server Updates (Gradients) & Trust Boundaries between Edge/Cloud \\ \hline

        Observability / Monitoring & High: Global logging \& SIEM integration & Low: Limited by device resources/privacy & Moderate: Obfuscated by Secure Aggregation & Variable: Fragmented monitoring logs \\ \hline

        Detection Capability & High (centralized monitoring) & Low & Medium–Low & Medium \\ \hline

        Typical Impact & System-wide & Persistent, device-specific & Gradual, hard to attribute & Cascading failures \\ \hline

        Scalability of Impact & Systemic: Single exploit affects all users & Isolated: Restricted to local instance & Collective: Degrades global model quality & Asymmetric: Cloud breach dwarfs edge leakage \\ \hline

        Trust Assumptions & Trust in Cloud Service Provider (CSP) & Local Sandbox \& Kernel Integrity & Trust in Aggregate Integrity (No-Sybil) & Mutual Authentication across Tiers \\ \hline

        Constraint Focus & Availability \& API Throughput & RAM, Battery, and Compute Power & Latency \& Communication Overhead & Orchestration \& Policy Uniformity \\ \hline
    \end{tabular}

    \label{tab:ragsecuritycomparision}
\end{table*}

\subsubsection{Federated RAG Threats}

Federated RAG combines retrieval-augmented generation with federated learning, enabling clients to collaborate without sharing raw data. However, this architecture creates a compound threat surface because retrieval encoders, local indices, model updates, and generation components may all become attack targets. Unlike centralized RAG, federated RAG operates under partial observability: no single party has complete visibility into client data, local retrieval behavior, or update provenance \cite{fedattack2}.

Model poisoning occurs when malicious clients submit manipulated updates to corrupt the shared retriever, ranker, or generator. In RAG, even subtle poisoning of embedding spaces can change which evidence is retrieved and therefore influence generation long after the attack round \cite{collusion2}. Gradient and update leakage represent the opposite risk: even when raw documents are not shared, model updates may reveal sensitive terms, topic distributions, retrieval preferences, or properties of local knowledge bases \cite{fedattack1}. These risks are amplified by non-IID data, small client populations, and partial participation.

Federated RAG is also vulnerable to Sybil and collusion attacks, where adversaries create fake clients or coordinate compromised clients to bias aggregation, amplify poisoning, or steer retrieval behavior \cite{collusion1}. Compound attacks may combine poisoned retrievers, context manipulation, gradient leakage, and Sybil behavior, making them harder to detect than isolated threats \cite{crosslayer1}. Defenses require retrieval-aware robust aggregation, client reputation, anomaly detection under non-IID data, secure aggregation with accountability, and mechanisms that jointly evaluate retrieval and generation behavior.

\subsubsection{Hybrid Edge--Cloud Threats}

Hybrid RAG partitions retrieval, context construction, and generation across local and cloud components. While this improves flexibility, it expands the attack surface across trust boundaries. Sensitive snippets, query embeddings, routing metadata, or partially processed context may cross from edge to cloud. Policy mismatches between local and cloud components can also lead to inconsistent enforcement of privacy and safety constraints. Therefore, hybrid RAG requires careful orchestration, mutual authentication, selective disclosure, encrypted communication, and policy-aware offloading.

\subsection{Cross-Paradigm Security Comparison}

Security risks in RAG systems are architecture-dependent. Centralized RAG benefits from global monitoring, logging, and rapid patching, but concentrates sensitive data and creates system-wide failure modes. On-device RAG reduces network exposure but is vulnerable to local tampering, side channels, and limited monitoring. Federated RAG avoids raw data sharing but introduces poisoning, update leakage, Sybil attacks, and partial observability. Hybrid RAG balances privacy and utility but creates additional risks at edge--cloud boundaries \cite{limitation1,fedeval2}.

\begin{table*}[hb]
\centering
\caption{Attack Types Across RAG Paradigms and Pipeline Stages}
\label{tab:attack_coverage}
\footnotesize
\begin{tabular}{|p{2.9cm}|p{2.3cm}|c|c|c|c|p{3.3cm}|}
\hline
\centering\textbf{Attack Type} & \centering\textbf{Pipeline Stage} 
& \centering\textbf{Centralized} 
& \centering\textbf{On-Device} 
& \centering\textbf{Federated} 
& \centering\textbf{Hybrid} 
& \centering\arraybackslash\textbf{Typical Signals} \\ 
\hline

Prompt / Instruction Injection 
& Query / Generation 
& \cmark & \cmark & \cmark & \cmark 
& Policy override, unsafe outputs \\ \hline

Knowledge Base Poisoning 
& Retrieval / Index 
& \cmark & \cmark & \cmark & \cmark 
& Biased ranking, misinformation \\ \hline

Membership Inference 
& Retrieval Output 
& \cmark & \cmark & \cmark & \cmark 
& Repeated hit-rate anomalies \\ \hline

Index Inference 
& Retrieval 
& \cmark & \cmark & \cmark & \cmark 
& Topic presence detection \\ \hline

Packing-Time Manipulation 
& Context Construction 
& \cmark & \cmark & \cmark & \cmark 
& Evidence truncation, dominance \\ \hline

Retriever Manipulation 
& Retrieval Model 
& \cmark & \cmark & \cmark & \cmark 
& Ranking instability \\ \hline

Model Extraction / Inversion 
& Generation 
& \cmark & \cmark & \cmark & \cmark 
& Parameter approximation \\ \hline

Gradient / Update Leakage 
& Training / Aggregation 
& \xmark & \xmark & \cmark & \cmark 
& Client data reconstruction \\ \hline

Sybil / Collusion Attacks 
& Training / Aggregation 
& \xmark & \xmark & \cmark & \cmark 
& Aggregation bias \\ \hline

Local Index Tampering 
& Storage 
& \xmark & \cmark & \xmark & \cmark 
& Integrity violations \\ \hline

\end{tabular}
\end{table*}

Table~\ref{tab:ragsecuritycomparision} summarizes security characteristics across deployment paradigms, while Table~\ref{tab:attack_coverage} maps major attacks to pipeline stages and deployment settings. The key implication is that RAG security cannot be treated as a single model-level problem. Effective protection requires architecture-aware defenses that account for where retrieval occurs, where context is constructed, where generation is executed, what metadata is exposed, and how trust is distributed across clients, devices, and cloud services.

\begin{figure}[ht]
\centering
\includegraphics[scale=0.4]{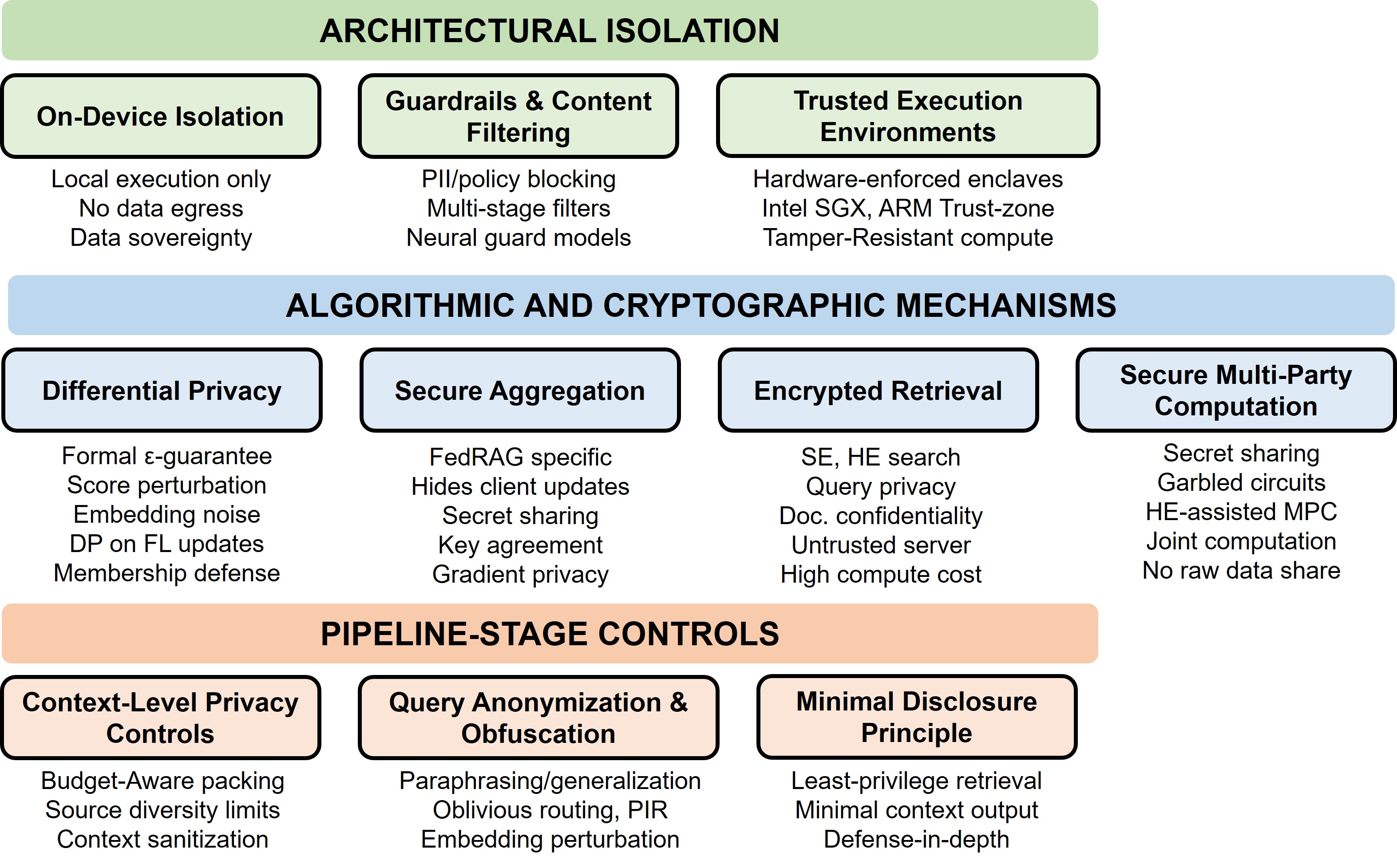}
\caption{Privacy-preserving techniques for RAG systems, organized across three mechanism classes: (a) architectural isolation, (b) algorithmic and cryptographic mechanisms, and (c) pipeline-stage controls.}
\label{fig:rag_privacy_tech}
\end{figure}

\section{Privacy-Preserving Techniques for Retrieval-Augmented Generation}
\label{sec:privacy}

The use of external knowledge sources in RAG introduces privacy risks through query logs, retrieval indices, embeddings, retrieved context, generated outputs, and federated updates. Unlike standalone LLMs, privacy leakage in RAG can occur at multiple stages: during query formulation, retrieval, context construction, generation, or collaborative training. Therefore, privacy-preserving RAG requires layered safeguards rather than a single mechanism. Fig.~\ref{fig:rag_privacy_tech} summarizes the main classes of privacy-preserving techniques considered in this section.

\subsection{Defense-in-Depth Framework for Trustworthy RAG}
\label{subsec:defense_in_depth}

\begin{figure*}[b]
\centering
\includegraphics[scale=0.4]{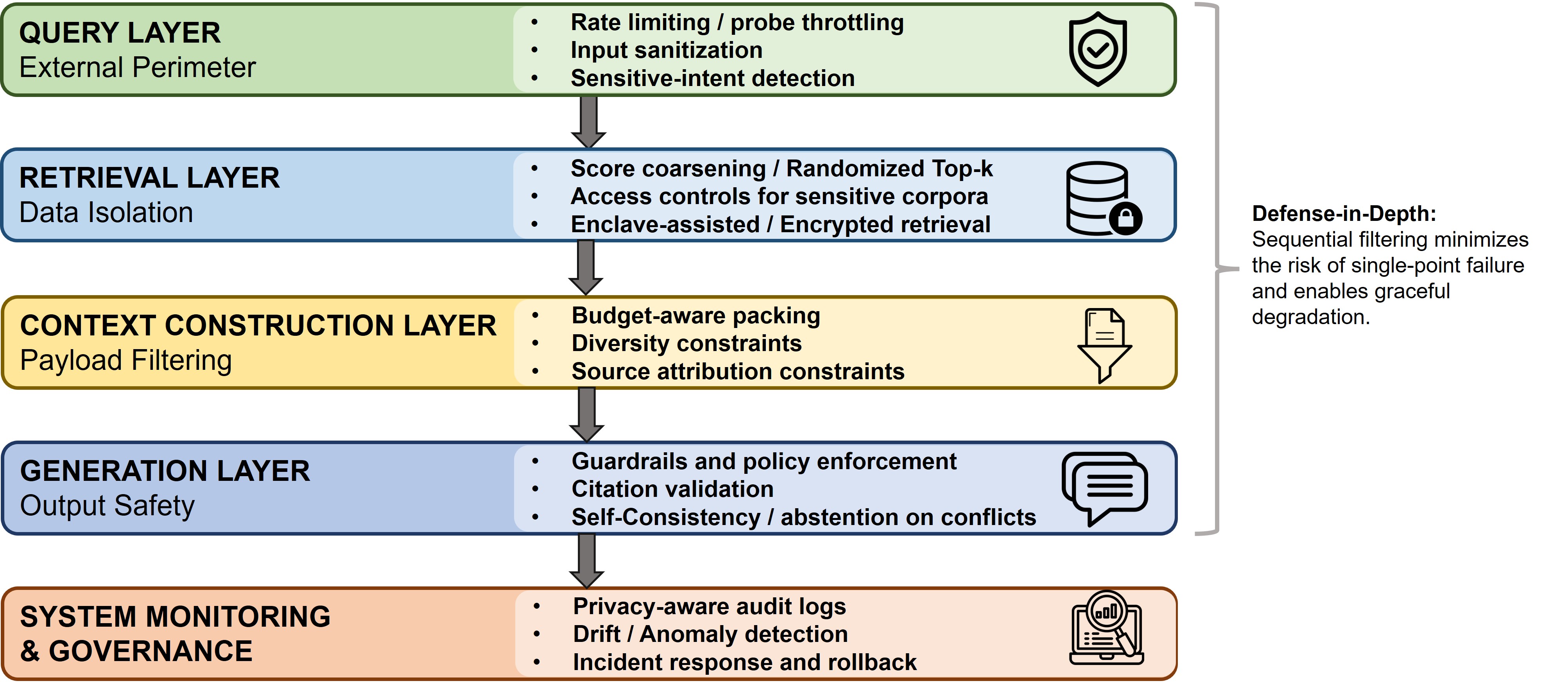}
\caption{Defense-in-depth framework for RAG systems. Privacy and security protections should be layered across the pipeline from query handling to retrieval, context construction, and generation, with monitoring and governance mechanisms providing operational resilience.}
\label{fig:rag_defense_in_depth}
\end{figure*}

A single privacy or security mechanism is insufficient to protect RAG systems, as leakage and manipulation can occur at multiple pipeline stages. As illustrated in Fig.~\ref{fig:rag_defense_in_depth}, trustworthy RAG requires a defense-in-depth framework with safeguards layered from query handling to retrieval, context construction, generation, and system-level monitoring. At the query stage, filtering, anonymization, and policy checks can reduce exposure of sensitive user intent. At the retrieval stage, access control, metadata suppression, score perturbation, encrypted retrieval, and anomaly detection can limit index leakage and adversarial probing. During context construction, redaction, source diversity, minimal-disclosure packing, and context sanitization help reduce unnecessary exposure of sensitive evidence. At the generation stage, guardrails, citation checking, refusal mechanisms, and faithfulness verification reduce privacy leakage and unsupported outputs. Finally, monitoring and governance mechanisms provide auditability, rate limiting, policy enforcement, and incident response across the full pipeline.

This layered view is particularly important for on-device, federated, and hybrid RAG deployments. On-device systems benefit from local isolation but still require tamper-resistant storage and lightweight verification. Federated RAG requires secure aggregation, robust client monitoring, and protection against update leakage or collusion. Hybrid edge--cloud systems require policy-aware offloading and secure coordination across trust boundaries. Thus, the defense-in-depth framework provides a unifying structure for comparing the privacy-preserving techniques discussed in the following subsections.

\subsection{Architectural and Pipeline-Level Controls}

Architectural isolation is the most direct way to reduce data exposure. In on-device or Micro-RAG systems, retrieval and, in some cases, generation are executed locally, ensuring that raw queries, private documents, and intermediate representations remain within the device boundary \cite{slmsurvey2}. This design supports data sovereignty and reduces server-side leakage, making it attractive for healthcare endpoints, enterprise devices, personal assistants, and privacy-sensitive IoT applications. However, local execution does not eliminate privacy risk: device compromise, malware, physical access, or side-channel leakage may still expose local indices or cached embeddings. Thus, on-device isolation should be combined with tamper-resistant storage, sandboxing, encryption, and lightweight integrity checks.

Guardrails and content filtering provide complementary protection by preventing sensitive or policy-violating information from entering or leaving the RAG pipeline. These controls can be applied during query preprocessing, document filtering, context construction, and output validation \cite{ragguardrails}. Rule-based filters, PII detectors, and neural guard models can reduce accidental disclosure of confidential records or proprietary content \cite{ragguardrails1}. However, guardrails remain vulnerable to paraphrasing, semantic obfuscation, and indirect leakage, and overly aggressive filtering can reduce retrieval recall and answer completeness \cite{ragguardrails2}. Hence, guardrails are best treated as one layer in a broader defense-in-depth strategy.

Pipeline-level controls specifically address privacy risks at the query and context-construction stages. Query anonymization, paraphrasing, generalization, perturbation, and oblivious routing can reduce exposure of sensitive user intent before retrieval occurs \cite{queryanoy}. Context-level privacy controls, such as source diversity constraints, redaction, pseudonymization, and minimal-disclosure context assembly, reduce the amount of sensitive information inserted into the prompt window \cite{badrag,sanitization,contextassembly}. These mechanisms are especially important in federated and hybrid RAG, where retrieved snippets or metadata may cross client or cloud boundaries.

\subsection{Algorithmic Privacy Mechanisms}

Differential Privacy (DP) provides formal protection by bounding how much any individual record can influence observable outputs \cite{dpsurvey}. In RAG, DP can be applied to retrieval scores, query or document embeddings, ranking outputs, and federated updates \cite{ragwithdp}. Such mechanisms reduce membership inference, index inference, and update leakage risks by limiting the precision with which an adversary can distinguish whether a document or client contribution is present.

Despite its formal guarantees, DP is difficult to apply in RAG without degrading utility. Small perturbations to similarity scores or embeddings can change the top-$k$ retrieval set, and this ranking change can propagate to generation quality \cite{dplimitaions}. In interactive systems, repeated queries also consume privacy budget over time. Therefore, DP-based RAG requires careful calibration of privacy parameters, query frequency, retrieval depth, and acceptable loss in answer quality.

In federated RAG, secure aggregation protects client updates by ensuring that the server observes only aggregated model updates rather than individual contributions \cite{dfrag}. This is useful for protecting retriever fine-tuning gradients, adapted embedding representations, and statistical sketches of local corpora \cite{secaggfedrag}. However, secure aggregation does not eliminate all leakage. Non-IID data, small client populations, partial participation, and collusion can weaken anonymity and make individual updates more inferable \cite{secaggfedrag1}. Moreover, secure aggregation increases communication and computation overhead, particularly when large retrieval encoders are collaboratively trained.

\subsection{Cryptographic and Hardware-Assisted Protection}

Encrypted retrieval aims to perform search over protected data without revealing plaintext queries or documents to the retrieval server \cite{encryptedret}. Searchable Symmetric Encryption (SSE) supports encrypted keyword search, while Homomorphic Encryption (HE) can support similarity computation over encrypted embeddings \cite{ssevshe}. These techniques provide strong confidentiality for cloud-hosted RAG, especially when the retrieval infrastructure should not access sensitive indexed content. However, encrypted vector search remains computationally expensive and less expressive than plaintext approximate nearest-neighbor retrieval, limiting its scalability for large, high-dimensional RAG indices \cite{remoterag}.

Trusted Execution Environments (TEEs) provide hardware-isolated regions for confidential computation, protecting code and data even from a privileged host operating system \cite{tees,tees1}. In RAG, TEEs can support confidential retrieval, tamper-resistant local indices, and verifiable cloud-side generation in hybrid deployments \cite{teebasedrag}. Compared with HE-based retrieval, TEEs offer lower overhead and support more flexible similarity operations. However, they remain vulnerable to hardware-specific side channels and are constrained by enclave memory limits, requiring careful system design for large-scale retrieval \cite{tsperf}.

Secure Multi-Party Computation (SMPC) enables multiple parties to jointly compute functions over private inputs without revealing those inputs to one another \cite{SMPC1}. In federated RAG, SMPC can support privacy-preserving aggregation, collaborative retrieval statistics, or secure score aggregation across clients \cite{SMPC2,smpcprotocols}. SMPC offers strong theoretical guarantees, but its computation and communication overhead are typically too high for end-to-end RAG over large retrieval encoders or high-dimensional vectors. Practical use is therefore more realistic for selected sensitive subcomponents rather than full-pipeline deployment \cite{smpcsurvey,smpcinrag}.

\begin{table}[ht]
\centering
\caption{Comparison of Privacy-Preserving Techniques Across RAG Paradigms}
\label{tab:privacy_comparison}
\footnotesize
\begin{tabular}{|p{2.5cm}|p{2.5cm}|p{2.2cm}|p{2cm}|p{1.5cm}|p{2.2cm}|}
\hline
\centering\textbf{Technique} & \centering\textbf{Primary Threat Mitigated} & \centering\textbf{Pipeline Stage} & \centering\textbf{Paradigm Fit} & \centering\textbf{Overhead} & \centering\arraybackslash\textbf{Deployment Maturity} \\ 
\hline

On-Device Isolation	& Query logging, server-side leakage &	All stages	& On-Device	& Low (network)	& High (production) \\ \hline
Guardrails / Filtering	& Prompt injection, PII leakage	& Query + Generation	& All	& Low	& High (production) \\ \hline
Differential Privacy (Retrieval) &	Membership inference, index inference	& Retrieval	& Centralized / Federated	& Moderate	& Medium (research) \\ \hline
Secure Aggregation (FL)	& Gradient / update leakage	& Training / Aggregation	& Federated	& High (comm.)	& Medium (production) \\ \hline
Encrypted Retrieval (SE/HE)	& Index confidentiality, query privacy	& Retrieval	& Centralized / Hybrid	& Very High	& Low (research) \\ \hline
Trusted Execution Environments	& Model extraction, side-channels, tampering	& All stages	& On-Device / Hybrid	& Low–Moderate	& Medium (emerging) \\ \hline
Secure Multi-Party Computation	& Cross-client inference, collusion	& Training / Retrieval &	Federated	& Very High	& Low (research) \\ \hline
Query Anonymization / Obfuscation	& Query-level inference, user profiling	& Query	& All	& Low	& Medium (research) \\ \hline
Context-Level Privacy Controls	& Packing-time leakage, truncation bias	& Context Construction	& All	& Low–Moderate	& Low (emerging) \\ \hline
Federated Encoder DP	& Encoder inversion, non-IID leakage	& Training (Encoder)	& Federated	& Moderate	& Low (research) \\ \hline
\end{tabular}
\end{table}

\subsection{Privacy--Utility Trade-off and Summary}

Privacy-preserving RAG mechanisms differ substantially in their guarantees, overhead, and deployment maturity. Architectural isolation and guardrails are practical and low-overhead, but they provide only limited formal guarantees. DP and secure aggregation offer stronger privacy for retrieval outputs and federated updates, but may reduce retrieval quality or increase communication cost. Encrypted retrieval and SMPC provide strong cryptographic protection, but remain expensive for large-scale semantic search. TEEs offer a practical middle ground, though they rely on hardware trust assumptions and remain vulnerable to side-channel attacks.

\begin{table*}[b]
\centering
\caption{Defense Mechanisms: Coverage and Trade-Offs in RAG Systems}
\label{tab:defense_coverage}
\footnotesize
\begin{tabular}{|l|l|p{2.1cm}|p{2.6cm}|p{1.6cm}|}
\hline
\centering\textbf{Defense Mechanism} 
& \centering\textbf{Primary Threat Mitigated} 
& \centering\textbf{Utility Impact} 
& \centering\textbf{Overhead} 
& \centering\arraybackslash\textbf{Deployment Complexity} \\ 
\hline

Guardrails / Filtering 
& Prompt injection, leakage 
& Low–Moderate 
& Low 
& Low \\ \hline

Score Coarsening / Randomized Top-$k$ 
& Membership, index inference 
& Moderate 
& Low 
& Moderate \\ \hline

Differential Privacy 
& Membership inference 
& Moderate–High 
& Low–Moderate 
& Moderate \\ \hline

Secure Aggregation 
& Gradient leakage 
& Low (accuracy) 
& High (communication) 
& High \\ \hline

Encrypted Retrieval 
& Index confidentiality 
& Low–Moderate 
& High (compute) 
& High \\ \hline

On-Device Isolation 
& Query logging, server leakage 
& Coverage constraints 
& Low (network) 
& Moderate \\ \hline

Anomaly Detection / Rate Limiting 
& Query probing, sybil attacks 
& Minimal 
& Low 
& Moderate \\ \hline

Hybrid Defense (Layered) 
& Multi-stage threats 
& Variable 
& Variable 
& High \\ \hline

\end{tabular}
\end{table*}

A central challenge is that most mechanisms protect only one stage of the RAG pipeline. For example, DP may protect retrieval scores while leaving packed context exposed; secure aggregation may protect federated updates while leaving local indices vulnerable; and encrypted retrieval may protect search but not generated outputs. Therefore, privacy-preserving RAG should be evaluated as a composable, end-to-end problem spanning query processing, retrieval, context construction, generation, and system monitoring. Table~\ref{tab:privacy_comparison} compares the major privacy-preserving techniques across paradigms, pipeline stages, overhead, and deployment maturity, while Table~\ref{tab:defense_coverage} summarizes their coverage and trade-offs.

\begin{figure}[ht]
\centering
\includegraphics[scale=0.60]{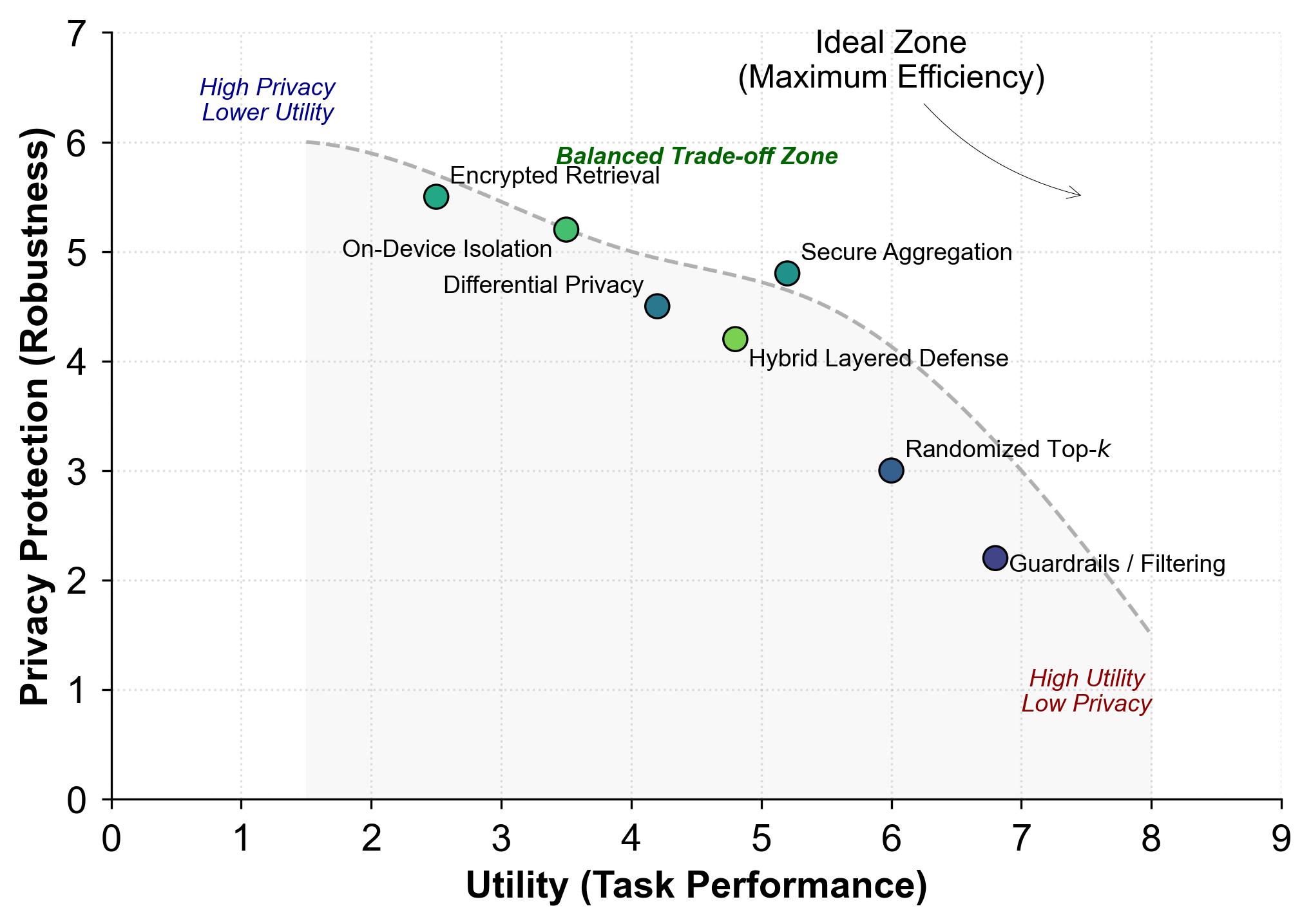}
\caption{Conceptual privacy--utility trade-off landscape for privacy-preserving RAG techniques. Architectural isolation and cryptographic methods offer stronger privacy guarantees but may reduce scalability or efficiency, while lightweight filtering mechanisms preserve utility with weaker privacy protection.}
\label{fig:rag_privacy_utility_tradeoff}
\end{figure}

The privacy--utility trade-off is unavoidable. Stronger privacy mechanisms often reduce retrieval accuracy, increase latency, consume communication bandwidth, or limit deployment scalability \cite{challenges1}. As illustrated in Fig.~\ref{fig:rag_privacy_utility_tradeoff}, the relative positions of privacy-preserving techniques should be interpreted qualitatively rather than as measured performance values. Future RAG systems require composable privacy frameworks that jointly account for leakage across retrieval, context construction, and generation, while maintaining acceptable accuracy and efficiency under centralized, on-device, federated, and hybrid deployments.

\section{Datasets, Benchmarks, and Evaluation Methodologies for RAG Security and Privacy}
\label{sec:datasets}

Evaluating RAG systems in privacy-sensitive and adversarial environments requires more than standard retrieval or generation accuracy. Metrics such as Recall@k, nDCG@k, Exact Match, and F1 remain useful for benign evaluation, but they do not capture adversarial retrieval, privacy leakage, context manipulation, federated robustness, or on-device efficiency. Therefore, security- and privacy-aware RAG evaluation must jointly consider retrieval quality, generation faithfulness, leakage risk, attack success, communication overhead, latency, and energy consumption.

\begin{table*}[htp]
\centering
\caption{Benchmark and Dataset Summary for RAG Security and Privacy Evaluation}
\label{tab:benchmark_dataset}
\footnotesize
\begin{tabular}{|p{2.5cm}|p{2cm}|p{1.5cm}|p{5cm}|p{1.5cm}|p{2.2cm}|}
\hline
\centering\textbf{Dataset / Benchmark} & \centering\textbf{Task Type} & \centering\textbf{Domain} & {Privacy / Security Relevance} & \centering\textbf{Paradigm Fit} & \centering\arraybackslash\textbf{Approx. Size} \\ 
\hline

MS MARCO \cite{msmacrods}	&	Passage Retrieval	&	Open Domain	&	Baseline retrieval; no privacy annotations; used to measure recall degradation under DP noise	&	Centralized	&	~8.8M passages \\ \hline
Natural Questions (NQ) \cite{naturalquestds}	&	Open-Domain QA	&	Wikipedia	&	Membership inference probing via QA hit-rate analysis; widely used for leakage baselines	&	Centralized / Federated	&	~300K pairs \\ \hline
BEIR Suite \cite{bierbenchmarks}	&	IR Heterogeneous	&	Multi-domain	&	Cross-domain retrieval robustness; evaluates generalization under distribution shift induced by poisoning	&	Centralized	&	18 datasets \\ \hline
TriviaQA \cite{triviaqa}	&	Open-Domain QA	&	Trivia / Web	&	Factual grounding evaluation; hallucination detection baseline under conflicting retrieval	&	Centralized	&	~95K pairs \\ \hline
HotpotQA \cite{hotpotqa}	&	Multi-hop QA	&	Wikipedia	&	Multi-document reasoning; evaluates context manipulation and evidence dominance in packing	&	Centralized	&	\textasciitilde113K pairs \\ \hline
CRAG \cite{cragbenchmark}	&	Factual Consistency QA	&	Multi-domain	&	Dynamic fact verification; evaluates faithfulness and hallucination under retrieved evidence	&	Centralized / On-Device	&	\textasciitilde4,400 QA pairs \\ \hline
DRAGOn \cite{dragon}	&	Dynamic Corpus IR	&	Clinical NLP	&	Temporal staleness and knowledge update evaluation; relevant for on-device index refresh	&	On-Device / Federated	&	Variable \\ \hline
RGB (Retrieval Benchmark for Generative models)	\cite{rgb} &	RAG Robustness	&	Open Domain	&	Evaluates noise robustness, counterfactual robustness, and information integration under adversarial retrieval	&	Centralized	&	\textasciitilde4,000 instances \\ \hline
RECALL	&	Counterfactual Robustness	&	Open Domain	&	Tests whether models over-rely on retrieved context vs. parametric knowledge; relevant to grounding dominance attacks	&	Centralized	&	\textasciitilde2,900 instances \\ \hline
PoisonedRAG \cite{poisoningattack2}	&	Retrieval Poisoning	&	Open Domain	&	Adversarial corpus injection benchmark; evaluates top-k displacement and misinformation injection attacks	&	Centralized / Federated	&	Custom adversarial \\ \hline
SafeRAG \cite{saferag}	&	Multi-Attack Evaluation	&	Open Domain	&	Covers noise, conflict, and injection attacks across the retrieval-to-generation pipeline; multi-stage evaluation	&	Centralized	&	\textasciitilde2,000 instances \\ \hline
BadRAG \cite{microrag2}	&	Backdoor / Vulnerability	&	Open Domain	&	Systematically probes RAG vulnerabilities including index tampering and model extraction vectors	&	On-Device / Centralized	&	Custom adversarial \\ \hline
PubMedQA \cite{pubmedqa}	&	Biomedical QA	&	Biomedical	&	Clinical RAG evaluation; sensitive domain requiring confidentiality; context leakage and hallucination risk	&	On-Device / Federated	&	\textasciitilde273K pairs \\ \hline
MedQA (USMLE) \cite{medqa}	&	Clinical QA	&	Medical	&	Multi-choice clinical reasoning; evaluates factual grounding under privacy-constrained retrieval in healthcare	&	On-Device / Federated	&	\textasciitilde61K pairs \\ \hline
HealthCareMagic \cite{healthcaremagic}	&	Patient Dialogue QA	&	Healthcare	&	Real patient–physician dialogues; high sensitivity; used for context leakage and privacy preservation evaluation	&	On-Device	&	\textasciitilde200K dialogues \\ \hline
CVE / NVD Repositories \cite{nvddatabase}	&	Security Knowledge IR	&	Cybersecurity	&	Sensitive vulnerability data; adversarial probing for index inference and unauthorized disclosure attacks	&	Centralized / On-Device	&	200K+ entries \\ \hline
LegalBench \cite{legalbench} 	&	Legal Reasoning QA	&	Legal / Regulatory	&	Legally privileged information; evaluates compliance-aware retrieval and selective disclosure mechanisms	&	Federated / Hybrid	&	\textasciitilde90 tasks \\ \hline
LEAF (FEMNIST, Shakespeare) \cite{leaf}	&	Federated Learning Eval	&	Multi-client	&	Non-IID data distribution evaluation; convergence and poisoning robustness under federated retrieval	&	Federated	&	Variable \\ \hline
FedQA (custom) \cite{fedqa}	&	Federated QA	&	Multi-domain	&	Federated RAG evaluation across heterogeneous clients; gradient leakage and Sybil attack baselines	&	Federated	&	Research-defined \\ \hline
PrivacyLens / PIE \cite{privacylens}	&	Privacy Inference Eval	&	Multi-domain	&	Membership inference and index inference evaluation; AUC-based leakage quantification under black-box settings	&	All Paradigms	&	Research-defined \\ \hline

\end{tabular}
\end{table*}

\begin{table*}[ht]
\centering
\caption{Evaluation Metrics Reference for RAG Security and Privacy}
\label{tab:evalmetrics}
\footnotesize
\begin{tabular}{|p{3cm}|p{1.5cm}|p{1.6cm}|p{4.1cm}|p{4.1cm}|}
\hline
\centering\textbf{Metric} & \centering\textbf{Category} & \centering\textbf{Pipeline Stage} & \centering\textbf{What It Measures} & \centering\arraybackslash\textbf{Key Limitation for RAG}  \\ 
\hline

Recall@k	&	Retrieval	&	Retrieval	&	Fraction of relevant docs in top-k	&	Does not penalise high-ranking adversarial docs \\ \hline
nDCG@k	&	Retrieval	&	Retrieval	&	Graded relevance with rank discounting	&	Assumes benign, static corpus; no poisoning model \\ \hline
MRR	&	Retrieval	&	Retrieval	&	Rank of first relevant result	&	Sensitive to ranking manipulation; no privacy signal \\ \hline
Exact Match	&	Generation	&	Generation	&	Binary accuracy vs. gold answer	&	Penalises valid paraphrases; ignores hallucination type \\ \hline
F1 / Token Overlap	&	Generation	&	Generation	&	Partial string overlap with gold	&	Cannot distinguish grounded from parametric hallucination \\ \hline
Faithfulness	&	Generation	&	Generation	&	Generated claims entailed by retrieved context	&	Entailment model can be fooled by adversarial context \\ \hline
Citation Accuracy	&	Generation	&	Generation	&	Correct attribution of claims to source docs	&	Can be gamed by adversarial document formatting \\ \hline
Refusal Correctness	&	Generation	&	Generation	&	Appropriate abstention on unanswerable queries	&	No standardised unanswerable query sets for RAG \\ \hline
Hallucination Rate	&	Generation	&	Generation	&	Proportion of responses with unsupported false claims	&	Requires ground-truth knowledge base for verification \\ \hline
MIA-AUC	&	Privacy	&	Retrieval	&	Adversary discriminates index members from non-members	&	AUC alone insufficient; report TPR@1\%,5\%,10\% FPR \\ \hline
TPR @ fixed FPR	&	Privacy	&	Retrieval	&	Attack success at operationally realistic false-alarm rate	&	Threshold must be standardised across studies \\ \hline
Index Inference Acc.	&	Privacy	&	Retrieval	&	Adversary reconstructs corpus topic structure via probing	&	No standardised benchmark; query budget varies \\ \hline
Output Privacy Leakage	&	Privacy	&	Generation	&	PII / sensitive content in generated responses	&	Requires annotated sensitive corpus; no standard tool yet \\ \hline
Attack Success Rate	&	Adversarial	&	Retrieval+Gen	&	Fraction of queries where attack achieves its goal	&	Attack definition varies across frameworks \\ \hline
Evidence Survival Rate	&	Adversarial	&	Context	&	Legitimate docs retained under adversarial packing	&	Not yet standardised across poisoning benchmarks \\ \hline
Byzantine Resistance	&	Federated	&	Aggregation	&	Quality retention under fraction of malicious clients	&	Non-IID data confounds poisoning signal \\ \hline
Gradient Leakage Score	&	Federated	&	Aggregation	&	Reconstruction fidelity of private data from gradients	&	Requires ground-truth private data; lab setting only \\ \hline
Comm. Overhead (bits/round)	&	Federated	&	Aggregation	&	Bits transmitted per federated training round	&	Trade-off with convergence rarely co-reported \\ \hline
End-to-End Latency (ms)	&	Efficiency	&	All	&	Wall-clock time from query to response	&	Device heterogeneity makes cross-study comparison unreliable \\ \hline
Energy per Query (mJ)	&	Efficiency	&	On-Device	&	Energy for one end-to-end inference on-device	&	Rarely reported; critical for battery-constrained RAG \\ \hline

\end{tabular}
\end{table*}

\subsection{Benchmark Categories}

Existing benchmarks can be grouped into five broad categories. First, retrieval-centric benchmarks such as MS MARCO \cite{msmacrods}, Natural Questions \cite{naturalquestds}, and BEIR \cite{bierbenchmarks} provide standard baselines for retrieval quality, but generally assume benign and static corpora. Second, RAG-specific and generation-oriented benchmarks such as CRAG \cite{cragbenchmark}, DRAGOn \cite{dragon}, RGB \cite{rgb}, and RECALL evaluate factual consistency, temporal staleness, noisy retrieval, and counterfactual robustness. Third, domain-specific datasets such as PubMedQA \cite{pubmedqa}, MedQA \cite{medqa}, HealthCareMagic \cite{healthcaremagic}, NVD/CVE \cite{nvddatabase}, and LegalBench \cite{legalbench} are important for evaluating RAG in sensitive settings where privacy, compliance, and attribution are operationally relevant. Fourth, adversarial benchmarks such as PoisonedRAG, SafeRAG, and BadRAG support controlled evaluation of retrieval poisoning, multi-stage attacks, and on-device vulnerabilities. Finally, federated and privacy-oriented datasets are needed to evaluate non-IID data, client heterogeneity, membership inference, index inference, and update leakage. Table~\ref{tab:benchmark_dataset} summarizes representative datasets and benchmarks across these categories.

\subsection{Evaluation Metrics}

RAG evaluation should cover multiple metric families rather than relying on a single accuracy score. Retrieval quality is typically measured using Recall@k, nDCG@k, and MRR, while generation quality is assessed using Exact Match, F1, faithfulness, citation accuracy, refusal correctness, and hallucination rate. Privacy evaluation requires leakage-oriented metrics such as MIA-AUC, TPR at fixed FPR, index inference accuracy, and output privacy leakage. Security evaluation requires attack success rate, evidence survival rate, robustness under poisoned or conflicting context, and degradation under adaptive probing. Federated and on-device RAG additionally require communication cost, Byzantine robustness, convergence stability, end-to-end latency, and energy per query. Table~\ref{tab:evalmetrics} consolidates these metrics and highlights their limitations in the RAG context.

\subsection{Privacy--Utility Trade-off}

Privacy-preserving mechanisms often introduce measurable utility or efficiency costs. Differential privacy can perturb retrieval rankings, encrypted retrieval can increase latency, and secure aggregation can increase communication overhead \cite{privacytradeoff1}. Hence, evaluation should report privacy and utility jointly rather than in isolation. A meaningful comparison should include retrieval utility, generation faithfulness, privacy leakage, attack robustness, and system overhead, enabling privacy--utility frontier analysis across deployment paradigms \cite{limitation42}. Such analysis is especially important for on-device and federated RAG, where privacy gains may be offset by reduced retrieval coverage, limited observability, or higher coordination cost.

\subsection{Limitations of Current Benchmarks}

Current RAG benchmarks remain insufficient for reproducible security and privacy evaluation. First, most benchmarks assume a static corpus, whereas production RAG systems continuously update their knowledge bases and may be vulnerable to update-time poisoning \cite{ragevallimit1}. Second, few datasets provide annotated sensitive subsets, making end-to-end leakage evaluation across retrieval, context construction, and generation difficult \cite{ragevallimit2}. Third, there is no standardized cross-paradigm protocol for comparing centralized, on-device, federated, and hybrid RAG under consistent query distributions, corpus scales, and attacker models \cite{ragevallimit3}. Fourth, on-device and federated deployments are under-instrumented because logging, activation access, and intermediate-state inspection are limited by resource or privacy constraints \cite{federatedrag2}. Addressing these gaps requires benchmark suites that jointly evaluate factuality, privacy leakage, adversarial robustness, latency, energy, and communication cost.

\section{Open Challenges and Future Directions}
\label{sec:open_challenges}

\begin{figure*}[ht]
\centering
\includegraphics[scale=0.3]{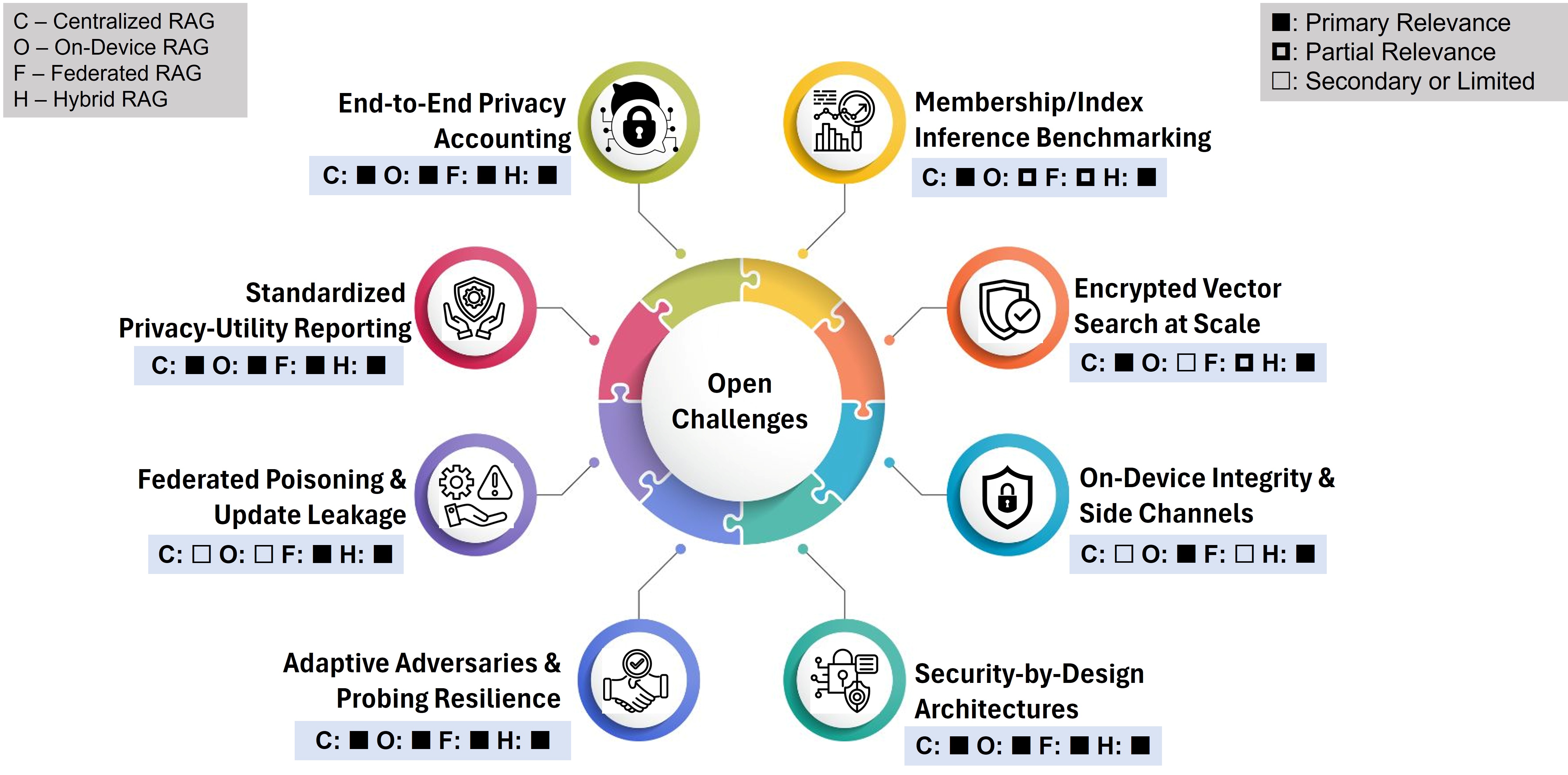}
\caption{Cross-mapping of open challenges in secure and privacy-preserving RAG to RAG taxonomies, with relevance levels (primary, partial, secondary) indicating where each challenge is most critical.}
\label{fig:openchallenges}
\end{figure*}

Despite rapid progress in RAG, deploying RAG systems in privacy-sensitive and adversarial environments remains challenging. This section highlights key open problems spanning trusted system design, robust retrieval, privacy guarantees, and evaluation standardization. Figure~\ref{fig:openchallenges} provides a paradigm-aware view of open challenges, highlighting how privacy and robustness priorities shift across centralized, on-device, federated, and hybrid RAG deployments.

\subsection{End-to-End Privacy Guarantees Across the RAG Pipeline}
Most privacy mechanisms target a single stage (e.g., secure aggregation for updates or differential privacy for retrieval scores), while end-to-end privacy remains poorly characterized. A major open challenge is developing composable privacy frameworks that jointly model leakage through queries, retrieval outputs, packed context, and generated responses. In particular, interactive usage patterns introduce cumulative privacy loss, motivating privacy accounting methods tailored to repeated retrieval and generation \cite{challenges2}.

\subsection{Benchmarking Membership and Index Inference at Scale}
Membership inference and index inference attacks are increasingly practical in black-box settings, yet standardized benchmarks remain limited. Future work should construct evaluation suites with annotated sensitive subsets, calibrated attacker capabilities (query budgets, feedback channels), and realistic enterprise-style corpora. Beyond AUC-based leakage metrics, systems should report operationally meaningful measures such as TPR at low FPR and privacy--utility frontiers under latency constraints \cite{limitation2}.

\subsection{Robust Retrieval Under Adaptive Adversaries}
Defenses such as score coarsening, randomized top-$k$, or filtering can reduce leakage, but adaptive attackers may exploit distribution shift, paraphrasing, or query reformulation to recover signal. Robust retrieval under adaptive probing remains an open problem, particularly when defenses must preserve utility for benign users. This motivates research into attack-aware retrieval policies, anomaly detection for probing behavior, and hybrid mechanisms that incorporate both statistical defenses and policy controls \cite{robustret}.

\subsection{Secure and Efficient Encrypted Retrieval for Vector Search}
While searchable encryption and homomorphic techniques provide strong confidentiality, integrating them with high-dimensional vector search remains computationally expensive. Key research directions include approximate encrypted similarity search, hardware-assisted secure enclaves for retrieval, and hybrid encryption strategies that protect only sensitive corpus segments. Practical solutions must address scalability, latency, and deployment complexity, especially for real-time RAG applications \cite{securevectorsearch}.

\begin{table}[hb]
\centering
\caption{Open Challenges and Future Research Directions in Secure and Privacy-Preserving RAG Systems}
\label{tab:open_challenges}
\footnotesize
\begin{tabular}{|p{3cm}|p{3.6cm}|p{3.8cm}|p{3.8cm}|}
\hline
\centering\textbf{Category} 
& \centering\textbf{Research Opportunity}
& \centering\textbf{Key Challenges}
& \centering\arraybackslash\textbf{Relevant Techniques / Directions} \\ 
\hline

End-to-End Privacy Guarantees 
& Develop composable privacy frameworks spanning retrieval, context construction, and generation 
& Cumulative privacy leakage across repeated queries; lack of unified accounting; cross-stage dependencies 
& Formal privacy accounting for RAG; pipeline-level DP; cross-stage leakage modeling \\ \hline

Membership and Index Inference 
& Standardized benchmarking and large-scale leakage evaluation 
& Absence of sensitive subsets; unrealistic attacker models; inconsistent metrics 
& Attack-driven evaluation suites; calibrated query budgets; privacy--utility frontier analysis \\ \hline

Adaptive Adversarial Retrieval 
& Retrieval policies robust to probing and reformulation attacks 
& Dynamic attacker behavior; balancing robustness and recall; avoiding utility degradation 
& Randomized ranking; anomaly detection; adaptive response shaping \\ \hline

Encrypted and Confidential Retrieval 
& Scalable encrypted similarity search for vector-based retrieval 
& High computational overhead; limited support for approximate nearest neighbors; latency constraints 
& Searchable encryption; homomorphic similarity; trusted execution environments \\ \hline

Federated RAG Robustness 
& Joint protection against poisoning and update leakage in decentralized retrieval 
& Non-IID data; collusion; gradient inversion; communication overhead 
& Secure aggregation with robustness guarantees; client reputation systems; Byzantine-resilient aggregation \\ \hline

On-Device RAG Integrity 
& Secure and tamper-resistant local retrieval pipelines 
& Device compromise; side-channel leakage; storage integrity 
& Hardware-backed isolation; lightweight attestation; encrypted local indices \\ \hline

Privacy--Utility Trade-off Quantification 
& Standardized reporting of joint privacy and performance metrics 
& Lack of common benchmarks; inconsistent leakage definitions; multi-objective optimization complexity 
& Unified evaluation protocols; trade-off visualization frameworks; cost-aware deployment models \\ \hline

Security-by-Design Architectures 
& Integrating layered defenses into RAG pipeline design 
& Retroactive defense integration; complexity across deployment paradigms 
& Defense-in-depth frameworks; least-privilege retrieval; minimal disclosure outputs \\ \hline

Context Construction Robustness 
& Formal guarantees for context allocation and truncation fairness 
& Budget-induced bias; evidence displacement; lack of formal models 
& Robust packing strategies; fairness-aware context selection; explainable attribution mechanisms \\ \hline

\end{tabular}
\end{table}

\subsection{Federated RAG: Coupled Risks in Learning and Retrieval}
Federated RAG inherits vulnerabilities from both federated learning and retrieval pipelines. Open challenges include preventing poisoning of shared retrievers, bounding leakage from updates under non-IID data, and handling sybil or collusion attacks. Moreover, federated systems introduce additional trust assumptions about aggregation servers and client participation. Future work should develop threat models and defenses that treat federated retrieval and generation as a coupled system rather than independent components \cite{pfedrag}.

\subsection{On-Device RAG: Local Privacy With New Attack Surfaces}
Micro-RAG reduces server-side exposure but introduces device-side risks, including local index tampering, side-channel leakage, and model extraction from compressed generators. Achieving strong privacy on-device without sacrificing coverage requires techniques for secure local storage, lightweight integrity checks, and resource-aware defenses. Establishing robust evaluation protocols for embedded and mobile deployments is also an open problem \cite{limitation6}.

\subsection{Trustworthy Evaluation Beyond Retrieval Accuracy}
RAG systems are often evaluated primarily on retrieval recall and downstream QA accuracy, which do not fully capture trustworthiness under adversarial or privacy-sensitive conditions. Future evaluations should incorporate faithfulness and attribution consistency, refusal correctness, leakage risk measures, and robustness under distribution shifts. Standardizing these metrics and reporting practices is essential for reproducible comparison across paradigms \cite{trustworthy}.

\subsection{Toward Architecture-Aware Security by Design}
A recurring theme across paradigms is that security and privacy failures frequently arise from architectural decisions (e.g., logging, caching, deterministic retrieval outputs, cross-device synchronization) rather than model weaknesses alone. Developing security-by-design principles for RAG, including least-privilege retrieval, minimal disclosure outputs, and defense-in-depth across pipeline stages, remains a central direction for building trusted and resilient RAG systems \cite{limitation81, limitation82}.

Table~\ref{tab:open_challenges} summarizes key research opportunities in secure and privacy-preserving RAG systems, highlighting unresolved technical challenges and promising directions across architectural, algorithmic, and evaluation dimensions. In summary, the transition from centralized to on-device and federated RAG represents a paradigm shift in how knowledge-grounded language models are deployed and trusted. Addressing the challenges outlined in this section will require interdisciplinary advances spanning machine learning, systems design, security, and privacy engineering. By identifying these open problems, this survey aims to guide future research toward scalable, trustworthy, and privacy-aware RAG systems that can operate effectively beyond centralized cloud environments.

\section{Summary and Conclusions}
\label{sec:conclusion}

Retrieval-Augmented Generation has emerged as a foundational paradigm for grounding language models in external knowledge, significantly improving factual accuracy and adaptability. However, the prevailing reliance on centralized RAG architectures limits deployability in privacy-sensitive, bandwidth-constrained, and resource-limited environments. This survey has examined the growing shift toward on-device and federated RAG systems, which seek to address these limitations by decentralizing retrieval and generation while preserving collaboration and performance.

We presented a unified taxonomy of RAG architectures spanning centralized, on-device (Micro-RAG), federated, and hybrid edge–cloud paradigms, highlighting their respective design trade-offs in terms of privacy, efficiency, scalability, and complexity. Beyond architectural considerations, we analyzed the expanded security and privacy threat surface introduced by RAG pipelines, including prompt-based attacks, retrieval poisoning, and federated adversarial behaviors. We further reviewed privacy-preserving mechanisms, ranging from on-device isolation and guardrails to differential privacy and secure aggregation, and discussed their applicability and limitations across various deployment settings.

Our paper also emphasized the importance of holistic evaluation, arguing that traditional accuracy-centric benchmarks are insufficient for decentralized RAG systems. Metrics capturing faithfulness, privacy leakage, communication cost, latency, and energy consumption are essential for assessing real-world viability, particularly in edge and federated contexts. The lack of standardized benchmarks and evaluation protocols remains a significant barrier to fair comparison and reproducible research.

Finally, we outlined key open challenges and research directions, including Micro-RAG design for constrained devices, communication-efficient federated RAG, handling non-IID and fragmented knowledge, trustworthy and verifiable generation, and security-aware system architectures. Addressing these challenges will require integrated advances across machine learning, systems engineering, security, and privacy.

In summary, on-device and federated RAG represent a critical evolution toward trustworthy, privacy-aware, and scalable language intelligence beyond the cloud. By consolidating current knowledge and identifying future opportunities, this survey aims to serve as a foundation for researchers and practitioners developing the next generation of decentralized RAG systems.

\bibliographystyle{unsrt}  
\bibliography{Bibliography}

\end{document}